\documentclass[aps,prd,floats,floatfix,preprintnumbers,superscriptaddress,nofootinbib]{revtex4-2}

\usepackage{graphicx} 
\usepackage{dcolumn}
\usepackage{amssymb}
\usepackage{amsmath}
\usepackage[hidelinks]{hyperref}
\usepackage{color}
\usepackage[utf8]{inputenc}
\usepackage{slashed}
\usepackage[normalem]{ulem}

\newcommand{\be}{\begin{equation}}
\newcommand{\ee}{\end{equation}}

\newcommand{\rmd}{\mathrm{d}}
\newcommand{\rme}{\mathrm{e}}
\newcommand{\rmi}{\mathrm{i}}
\newcommand{\calA}{\mathcal{A}}

\newcommand{\calJ}{\mathcal{J}}
\newcommand{\calP}{\mathcal{P}}
\newcommand{\calR}{\mathcal{R}}
\newcommand{\calM}{\mathcal{M}}
\newcommand{\calN}{\mathcal{N}}

\newcommand{\calD}{\mathcal{D}}
\newcommand{\calS}{\mathcal{S}}
\newcommand{\calO}{\mathcal{O}}

\newcommand{\rp}{\boldsymbol{r}_{\perp}}
\newcommand{\rop}{\boldsymbol{r}_{1\perp}}
\newcommand{\rtp}{\boldsymbol{r}_{2\perp}}

\newcommand{\bp}{\boldsymbol{b}_{\perp}}
\newcommand{\bop}{\boldsymbol{b}_{1\perp}}
\newcommand{\btp}{\boldsymbol{b}_{2\perp}}
\newcommand{\lp}{\boldsymbol{l}_{\perp}}

\newcommand{\xp}{\boldsymbol{x}_{\perp}}
\newcommand{\yp}{\boldsymbol{y}_{\perp}}
\newcommand{\zp}{\boldsymbol{z}_{\perp}}

\newcommand{\pp}{\boldsymbol{p}_{\perp}}

\newcommand{\kp}{\boldsymbol{k}_{\perp}}

\newcommand{\delp}{\boldsymbol{\Delta}_{\perp}}
\newcommand{\dlp}{\boldsymbol{\delta}_{\perp}}

\newcommand{\pd}{\partial}

\begin{document}
\date{\today}
\preprint{ZTF-EP-23-03}

\title{Exclusive $\eta_c$ production from small-$x$ evolved Odderon at a electron-ion collider}
\author{Sanjin Beni\' c}
\affiliation{Department of Physics, Faculty of Science, University of Zagreb, Bijenička c. 32, 10000 Zagreb, Croatia}
\author{Davor Horvati\' c}
\affiliation{Department of Physics, Faculty of Science, University of Zagreb, Bijenička c. 32, 10000 Zagreb, Croatia}
\author{Abhiram Kaushik}
\affiliation{Department of Physics, Faculty of Science, University of Zagreb, Bijenička c. 32, 10000 Zagreb, Croatia}
\author{Eric Andreas Vivoda}
\affiliation{Department of Physics, Faculty of Science, University of Zagreb, Bijenička c. 32, 10000 Zagreb, Croatia}
\begin{abstract} 
We compute exclusive $\eta_c$ production in high energy electron-nucleon and electron-nucleus collisions that is sensitive to the Odderon. In perturbative QCD the Odderon is a $C$-odd color singlet consisting of at least three $t$-channel gluons exchanged with the target.
By using the Color Glass Condensate effective theory our result describes the Odderon exchange at the high collision energies that would be reached at a future electron-ion collider. The Odderon distribution is evolved to small-$x$ using the Balitsky-Kovchegov evolution equation with running coupling corrections.
We find that while at low momentum transfers $t$ the cross section off a proton is dominated by the Primakoff process, the Odderon becomes relevant at larger momentum transfers of $|t|\geq1.5$ GeV$^2$. We point that the Odderon could also be extracted at low-$t$ using neutron targets since the Primakoff component is strongly suppressed. In the case of nuclear targets,  the Odderon cross section becomes enhanced thanks to the mass number of the nuclear target. The gluon saturation effect induces a shift in the diffractive pattern with respect to the Primakoff process that could be used as a signal for the Odderon.
\end{abstract}


\maketitle

\section{Introduction and motivation}

The Odderon was suggested 50 years ago \cite{Lukaszuk:1973nt,Ewerz:2003xi} as the $C$-odd ($C = -1$) partner of the $C-$even ($C = +1$) Pomeron in mediating a $t$-channel colorless exchange in elastic hadronic cross sections. The original idea \cite{Donnachie:1985iz} to measure the Odderon through a difference in $pp$ vs $p\bar{p}$ elastic cross sections brought much excitement recently \cite{TOTEM:2020zzr} thanks to the precise $pp$ measurement by the TOTEM collaboration \cite{TOTEM:2018psk} at the collision energies close to the $p\bar{p}$ D0 Tevatron data \cite{D0:2012erd}. On the other hand, considering elastic hadronic cross sections makes it difficult to understand the Odderon in the context of perturbative QCD.

As opposed to $pp$ collisions, $ep$ collisions provide a cleaner environment to extract the Odderon, particularly in the exclusive production of particles with a fixed $C$-parity. A prominent example here is the $\eta_c$ production \cite{Czyzewski:1996bv,Engel:1997cga,Bartels:2001hw,Ma:2003py,Goncalves:2012cy,Goncalves:2015hra,Goncalves:2018yxc,Harland-Lang:2018ytk,Dumitru:2019qec,Babiarz:2023cac} where the heavy charm quarks ensure that the process is sensitive to the gluons in the target. With the $C$-parity of $\eta_c$ being $C = +1$ and that of the emitted photon being $C= -1$, the amplitude becomes directly proportional to the Odderon. $\eta_c$ plays a role analogous to the $J/\psi$ production in case of the Pomeron. Unlike $J/\psi$, which has been extensively measured at HERA, there is no measurement of exclusive $\eta_c$ production so far. This would hopefully change with the high luminosities feasible at the upcoming Electron-Ion Colliders (EIC) \cite{Accardi:2012qut,LHeCStudyGroup:2012zhm,Anderle:2021wcy} (or even with the LHC in the ultra-peripheral mode \cite{Bertulani:2005ru}) and is therefore a motivation for our work. 

The high collision energies that will be reached at the EIC can offer unique insights into the small-$x$ component of the target wavefunction ($x$ represents the parton momentum fraction) where the gluon density is large according to the effective theory of the Color Glass Condensate (CGC) \cite{Iancu:2003xm,Gelis:2010nm,Kovchegov:2012mbw,Blaizot:2016qgz}. Within the framework of CGC, the Odderon is the imaginary part of the dipole distribution \cite{Kovchegov:2003dm,Hatta:2005as}
\be
\calO(\xp,\yp) \equiv \frac{1}{2\rmi N_c}{\rm tr} \left\langle V^\dag(\xp)V(\yp) - V^\dag(\yp)V(\xp)\right\rangle\,,
\label{eq:odder}
\ee
with the trace taken in the fundamental representation. The Wilson line $V(\xp)$ is defined in Sec.~\ref{sec:cs} below, in Eq.~\eqref{eq:Wilson}. The small-$x$ evolution of the Odderon is given by the imaginary part of the Balitsky-Kovchegov (BK) equation for the dipole \cite{Kovchegov:2003dm,Hatta:2005as,Motyka:2005ep}. Indeed, one of our main goals is to numerically solve the coupled Pomeron-Odderon BK system for the case of the proton and for nuclear targets. Whereas in the linear regime the Odderon and the Pomeron evolution is independent, the non-linearity of the BK equations alters the Odderon significantly when the dipole size is of the order of the inverse of the saturation scale $Q_S$ \cite{Kovchegov:2003dm,Hatta:2005as,Motyka:2005ep,Lappi:2016gqe,Yao:2018vcg}.

From a theoretical perspective, the difficulty in computing the $\eta_c$ cross section comes from the uncertainty in the magnitude of the Odderon. While earlier works on $\eta_c$ production \cite{Engel:1997cga,Czyzewski:1996bv,Bartels:2001hw,Ma:2003py} suggest a differential photo-production cross section in the range of $10^2$ pb/GeV$^2$, more recent computations \cite{Dumitru:2019qec} indicate that the cross section would be somewhat smaller -- of the order of 10$^2$ fb/GeV$^2$, and therefore overshadowed by the large background due to the Primakoff process in the low-$|t|$ region. This could be circumvented by considering instead neutron targets for which the low-$|t|$ Coulomb tail is absent allowing the Odderon to be probed even at low-$|t|$. These studies so far have focused on the Odderon in the dilute regime where $x$ is moderate and gluon density is not too large. Theoretical computations of the $\eta_c$ cross sections in the case of a dense proton or a nuclear target are so far unexplored and constitute another of our motivations.

In Sec.~\ref{sec:cs} we undertake the computation of the amplitude for $\eta_c$ production in the CGC formalism.
In Sec.~\ref{sec:bk} we solve the coupled Pomeron-Odderon BK system numerically using the kernel with running-coupling corrections and in the approximation where the impact parameter is treated as an external parameter \cite{Lappi:2013zma}. For the Pomeron initial condition we are using a fit to the HERA data (supplemented by optical Glauber in case of nuclei) \cite{Lappi:2013zma}. For the Odderon initial condition in case of nucleon targets we consider a recent computation in the light-cone non-perturbative quark model by Dumitru, M\"antysaari and Paatelainen \cite{Dumitru:2022ooz}. In case of nuclear targets we rely on a small-$x$ action with a cubic term in the random color sources \cite{Jeon:2005cf}. Sec.~\ref{sec:num} is devoted to the numerical results for the exclusive $\eta_c$ photo-production for the proton and the nuclear targets. Our main findings, laid out in the concluding Sec.~\ref{sec:conc}, are as follows. Probing the Odderon using proton targets requires rather high momentum transfers $|t| \gtrsim 1-3$ GeV$^2$ to access the region where the Primakoff background is subdominant. In case of neutron targets we find the Primakoff contribution to be negligible, allowing in principle, the extraction of the Odderon even at low-$|t|$. For nuclear targets the Odderon (Primakoff) cross section becomes enhanced roughly as $\sim A^2$ ($\sim Z^2$), where $A$ ($Z$) stand for the mass (atomic) number. The diffractive pattern in the Odderon cross section gets shifted by a few percent in comparison to the Primakoff cross section. This could serve as a distinctive signature of the Odderon.

\section{The cross section for exclusive $\eta_c$ production in the CGC framework}
\label{sec:cs}

The amplitude and the cross section for exclusive $\eta_c$ production $\gamma^*(q) p(P) \to \eta_c(\Delta) p(P')$ has been recently computed using light-cone wave functions at leading twist for the Odderon in \cite{Dumitru:2019qec}. For earlier works see \cite{Czyzewski:1996bv,Bartels:2001hw}. While some of the results from \cite{Dumitru:2019qec} carry over to our computations we find it worthwhile to quickly go over the derivation of the amplitude starting from the CGC framework \cite{Iancu:2003xm,Jalilian-Marian:2005ccm,Gelis:2010nm,Kovchegov:2012mbw,Blaizot:2016qgz} in momentum space and also taking into account the all-order multiple scatterings on a target, that is a dense proton or a nucleus. The cross section is computed in the frame where the target is moving along the light-cone minus coordinate, so that its momenta is $P^\mu = (P^+,0,\boldsymbol{0}_\perp)$, and that of the virtual photon $q^\mu = (q^+,q^-,\boldsymbol{0}_\perp)$\footnote{We are using light-cone variables: for a general vector $x^\mu = (x^0,x^1,x^2,x^3) = (x^+,x^-,\boldsymbol{x}_\perp)$ we have $x^{\pm} = (x^0 \pm x^3)/\sqrt{2}$.  Furthermore, we adhere to the following conventions: $\epsilon_{0123} = +1 = - \epsilon^{0123} = \epsilon^{+-12}$ with $\gamma^5 = \rmi\gamma^0 \gamma^1 \gamma^2 \gamma^3$.}. 
As for the kinematic variables of the process we denote with $t$ the momentum transfer
\be
t \equiv (P - P')^2 = -\frac{\delp^2}{1-x}\,,
\ee
where $x$ is the momentum fraction carried by the exchanged Odderon
\be
x \equiv \frac{(P - P')\cdot q}{P\cdot q} = \frac{Q^2 + M_{\calP}^2 - t}{W^2 + Q^2}\,,
\label{eq:xP}
\ee 
and $W^2 = (q + P)^2$ is the invariant mass of the $\gamma^*$-target system. We have $q^2 = - Q^2$ as the photon virtuality, $P^2 = P'^2 = 0$ and $\Delta^2 = M^2_\calP$ is the squared mass of the produced $\eta_c$ particle.

\subsection{The Odderon contribution}
\label{sec:odder}

\begin{figure}
  \begin{center}
  \includegraphics[scale = 0.4]{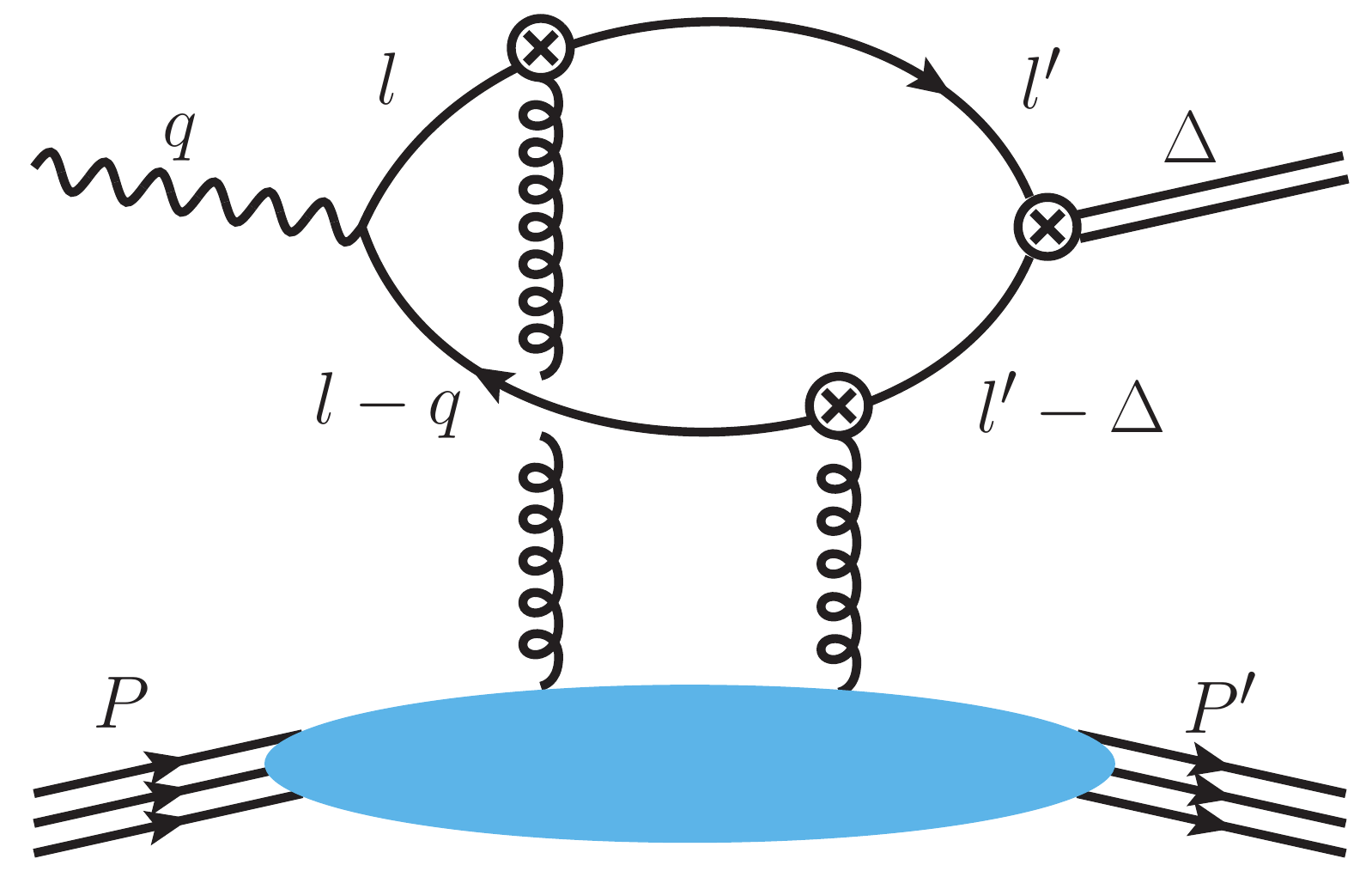}
  \end{center}
  \caption{Feynman diagram for exclusive $\eta_c$ production amplitude$\gamma^*(q) p(P) \to \eta_c(\Delta) p(P')$. The crosses where the vertical gluon lines attach to the $q\bar{q}$ state represent the effective CGC vertex \eqref{eq:cgcvert}.}
  \label{fig:etac}
\end{figure}

The amplitude for exclusive $\eta_c$ production can be written in complete analogy to that for $J/\psi$ production -- for a very clear recent exposition see for example \cite{Mantysaari:2020lhf}. We follow closely the notation used in \cite{Mantysaari:2020lhf} and write the CGC amplitude for $\eta_c$ production as
\be
\calS_\lambda = eq_c\int_{ll'}{\rm Tr}\left[S(l)\slashed{\epsilon}(\lambda,q)S(l-q)\tau(l-q,l'-\Delta)S(l'-\Delta)(\rmi \gamma_5)S(l')\tau(l',l)\right]\,,
\ee
where $q_c = 2/3$ is the charge of the charm quark in units of $e = \sqrt{4\pi \alpha}$, $\alpha = 1/137$ with $l$ and $l'$ representing the charm quark momenta as in Fig.~\ref{fig:etac}. We work in the $A^- = 0$ gauge where the virtual photon polarization vector $\epsilon^{\mu}(\lambda,q)$ is given as $\epsilon^{\mu}(0,q) = (Q/q^-,0,\boldsymbol{0}_\perp)$, $\epsilon^{\mu}(\lambda =\pm 1,q) = (0,0,\boldsymbol{\epsilon}^{\lambda}_\perp) = (0,0,1,\lambda \rmi)/\sqrt{2}$ and
\be
S(l) = \frac{\rmi(\slashed{l} + m_c)}{l^2 - m_c^2 + \rmi \epsilon}\,,
\ee
is the charm quark propagator with mass $m_c$. 
We use $(\rmi\gamma_5)$ as the Dirac structure for the vertex for $\eta_c$ production \cite{Czyzewski:1996bv,Dumitru:2019qec}, for the moment treating the $\eta_c$ wave function in perturbation theory.
For the phenomenological computation this will be replaced with a non-perturbative model $\eta_c$ light-cone wave function \cite{Kowalski:2006hc,Dumitru:2019qec}, see Eq.~\eqref{eq:boosted} below.
Inserting the effective CGC vertex \cite{McLerran:1998nk,Balitsky:2001mr} (see also \cite{Blaizot:2004wv}), 
\be
\tau(p,p') = (2\pi)\delta(p^- - p'^-)\gamma^- {\rm sgn}(p^-)\int_{\zp} \rme^{-\rmi (\pp - \pp')\cdot \zp} V^{{\rm sgn}(p^-)}(\zp)\,,
\label{eq:cgcvert}
\ee
where
\be
V(\zp) = \calP \exp\left[-\rmi g \int_{-\infty}^\infty \rmd y^- \frac{1}{\boldsymbol{\pd}_\perp^2} \rho^a(y^-,\zp)t^a \right]\,,
\label{eq:Wilson}
\ee
with $\rho^a(y^-,\zp)$ being the classical color source in the target, the amplitude becomes
\be
\begin{split}
\calS_\lambda &= - e q_c(2\pi)\delta(q^- - \Delta^-)\int_{ll'}(2\pi)\delta(l^- - l'^-)\theta(l^-)\theta(q^- - l^-)\int_{\xp\yp}\rme^{-\rmi (\lp' - \lp)\cdot \xp}\rme^{-\rmi(\lp - \lp' + \delp)\cdot \yp}\\
&\times{\rm tr}\left[V(\xp)V^\dag(\yp)\right]{\rm tr}\left[S(l)\slashed{\epsilon}(\lambda,q)S(l-q)\gamma^-S(l'-\Delta)(\rmi \gamma_5)S(l')\gamma^-\right]\,,
\end{split}
\label{eq:smatrix}
\ee
where the $\theta$-functions are dictated by the singularities of the quark propagators in the complex $l^+$ and $l'^+$ plane.

We can conveniently project out the Odderon by considering a diagram with the fermion flow in the opposite direction. 
Of course, with appropriate change of integration variables this simply gives back \eqref{eq:smatrix}. Utilizing instead $C$-parity transformation only on the Dirac part the resulting trace has an opposite sign to \eqref{eq:smatrix}. Combining the two contributions we come up with the (color averaged) amplitude as
\be
\left\langle\calS_\lambda\right\rangle = -\left\langle\calM_\lambda\right\rangle(2\pi)\delta(q^- - \Delta^-)\,,
\ee
where the amplitude $\langle \calM_\lambda \rangle$ is
\be
\begin{split}
\langle\calM_\lambda\rangle &= e q_c\int_{\rp}\int_{ll'}(2\pi)\delta(l^- - l'^-)\theta(l^-)\theta(q^- - l^-)\rme^{-\rmi (\lp' - \lp - \frac{1}{2}\delp)\cdot \rp}\\
&\times(-\rmi N_c)\calO(\rp,\delp){\rm tr}\left[S(l)\slashed{\epsilon}(\lambda,q)S(l-q)\gamma^-S(l'-\Delta)(\rmi \gamma_5)S(l')\gamma^-\right]\,.
\end{split}
\ee
with the Odderon distribution
\be
\calO(\rp,\delp) = \int_{\bp} \rme^{-\rmi \delp\cdot\bp}\calO(\rp,\bp)\,,
\label{eq:oddmom}
\ee
explicitly projected out. We have used $\rp = \xp - \yp$, $\bp = (\xp + \yp)/2$ and $\calO(\rp,\bp) \equiv \calO\left(\bp + \frac{\rp}{2},\bp - \frac{\rp}{2}\right)$ for short.

It is convenient to further separate out the Odderon distribution from the rest as
\be
\langle\calM_\lambda\rangle = (2q^-) \rmi N_c \int_{\rp} \calO(\rp,\delp)\calA_\lambda(\rp,\delp)\,,
\label{eq:mOdderon}
\ee
where the reduced amplitude $\calA_\lambda(\rp,\delp)$ (after light-cone $l^+$ and $l'^+$ integrals) is given as
\be
\calA_\lambda(\rp,\delp) = e q_c \int_z \int_{\lp \lp'}\frac{\rme^{\rmi (\lp - \lp' + \frac{1}{2}\delp)\cdot\rp} A_\lambda(l,l')}{\left(\lp^2 + \varepsilon^2\right)\left((\lp' - z \delp)^2 + \varepsilon'^2\right)}\,,
\label{eq:A1}
\ee
and
\be
A_\lambda(l,l') = \frac{\rmi}{(2q^-)^2}{\rm tr}\left[(\slashed{l}+m_c)\slashed{\epsilon}(\lambda,q)(\slashed{l}-\slashed{q} + m_c)\gamma^-(\slashed{l}' - \slashed{\Delta} + m_c)\gamma_5(\slashed{l}' + m_c)\gamma^-\right]\,.
\label{eq:Adir}
\ee
We have used the following abbreviations: $z \equiv l'^- /q^-$, $\varepsilon \equiv \sqrt{m_c^2 + z\bar{z}Q^2}$ and $\varepsilon' \equiv \sqrt{m_c^2 + z\bar{z}Q'^2}$\footnote{Formally we have $Q'^2 = -M_\calP^2$, and so the perturbative wave-function would become singular for time-like momenta. However, this becomes irrelevant in practice as we are replacing the perturbative wave function with a model, see \eqref{eq:Aadr}.} with $\bar{z} = 1-z$ and
\be
\int_z \equiv \int\frac{\rmd z}{4\pi}\,.
\ee
Computing the Dirac trace in \eqref{eq:Adir} we find
\be
A_\lambda(l,l') = 2m_c \epsilon^{+-ij}\epsilon^\lambda_{\perp i}(l_\perp - l'_\perp + z\Delta_\perp)_j\,.
\label{eq:trace}
\ee
The result \eqref{eq:trace} is proportional to $m_c$ because the Dirac trace contains 4 vertices and 3 fermion propagators in addition to $\gamma_5$. Intuitively, when the photon splits into a $q\bar{q}$ pair their spins are aligned, and not flipped by the eikonal interaction with the target. In order for the $q\bar{q}$ to combine into a spinless meson after the collision, we need a spin flip and this is provided by $m_c$. As another consequence of the eikonal interaction, we find the longitudinal photon $\lambda = 0$ decouples, as already noticed in \cite{Czyzewski:1996bv,Dumitru:2019qec} and in a related process in \cite{Boussarie:2019vmk}.

After computing the $\lp$ and $\lp'$ integrals we find
\be
\begin{split}
\calA_\lambda(\rp,\delp)
& = e q_c \lambda\rme^{\rmi \lambda \phi_r}\int_z \rme^{-\rmi \dlp\cdot\rp} (-1)\frac{\sqrt{2} m_c}{2\pi}\frac{1}{z\bar{z}} \left[K_0(\varepsilon r_\perp) \partial_{r_\perp}\phi_\calP(z,r_\perp) - \varepsilon K_1(\varepsilon r_\perp)\phi_\calP(z,r_\perp)\right]\\
&\equiv e q_c \lambda\rme^{\rmi \lambda \phi_r} \int_z \rme^{-\rmi \dlp\cdot\rp} \calA(r_\perp)\,,
\end{split}
\label{eq:Aadr}
\ee
where $\dlp \equiv \frac{1}{2}(z-\bar{z})\delp$ is the off-forward phase \cite{Hatta:2017cte} and we have separated out the $\lambda$ and $\delp$ independent part of the reduced amplitude as $\calA(r_\perp)$. 
We have also introduced the standard replacement \cite{Kowalski:2006hc} $K_0(\varepsilon' r_\perp)/(2\pi) \to \phi_\calP(z,r_\perp)$ to write the amplitude in terms of the $\eta_c$ meson light-cone wave function $\phi_\calP(z,r_\perp)$ \cite{Kowalski:2006hc,Dumitru:2019qec}.
In the numerical computations we are using a ``Boosted Gaussian" ansatz from \cite{Dumitru:2019qec}
\be
\phi_\calP(z,r_\perp) = \calN_P z\bar{z} \exp\left(-\frac{m_c^2\calR_\calP^2}{8z\bar{z}}-\frac{2z\bar{z}r_\perp^2}{\calR_\calP^2}+ \frac{1}{2}m_c^2 \calR_\calP^2\right)\,,
\label{eq:boosted}
\ee
with $\calN_\calP = 0.547$, $\calR^2_\calP = 2.48$ GeV$^{-2}$ and $m_c = 1.4$ GeV \cite{Dumitru:2019qec}.
The integrand in \eqref{eq:Aadr} can be understood as a $\gamma^*-\eta_c$ wave function overlap. Our result differs from (48) in \cite{Dumitru:2019qec} obtained using light-cone wave function approach by a relative sign between the two terms in the square bracket. Ref.~\cite{Dumitru:2019qec} uses the $\gamma^*$ wave function from \cite{Kowalski:2006hc}, however this is known to be incorrect, see e.~g. \cite{Lappi:2020ufv}. Using instead the $\gamma^*$ wave function from \cite{Dosch:1996ss,Lappi:2020ufv} we have explicitly confirmed the result in \eqref{eq:Aadr}.

It is useful to parametrize the Odderon distribution by the Fourier series
\be
\calO(\rp,\bp) = 2 \sum_{k = 0}^\infty\calO_{2k+1}(r_\perp,b_\perp)\cos((2k+1)\phi_{rb})\,,
\ee
where $\phi_{rb} \equiv \phi_r - \phi_b$. We calculate $\calO_{2k+1}(r_\perp,b_\perp)$ as
\be
\calO_{2k+1}(r_\perp,b_\perp) = \frac{1}{2\pi}\int_0^{2\pi} \rmd \phi_{rb} \calO(\rp,\bp) \cos((2k+1)\phi_{rb})\,.
\ee
We will consider its Fourier transform \eqref{eq:oddmom} $\calO(\rp,\delp)$ and expand it in Fourier series
\be
\calO(\rp,\delp) =  2 \sum_{k = 0}^\infty\calO_{2k+1}(r_\perp,\Delta_\perp)\cos((2k+1)\phi_{r\Delta})\,,
\ee
where
\be
\calO_{2k+1}(r_\perp,\Delta_\perp) = - 2\pi \rmi (-1)^k \int_0^\infty b_\perp \rmd b_\perp \calO_{2k+1}(r_\perp,b_\perp)J_{2k+1}(\Delta_\perp b_\perp)\,.
\label{eq:oddft}
\ee
With this parametrization the amplitude \eqref{eq:mOdderon} can be found in the following form
\be
\langle\calM_\lambda \rangle = q^-\lambda\rme^{\rmi \lambda \phi_\Delta}\langle\calM\rangle\,,
\label{eq:Mlambda}
\ee
where we have conveniently factored out the polarization independent amplitude $\langle \calM \rangle$ as
\be
\langle \calM\rangle = 8\pi \rmi e q_c N_c \sum_{k = 0}^\infty (-1)^k\int_z \int_0^\infty r_\perp \rmd r_\perp \calO_{2k+1}(r_\perp,\Delta_\perp) \calA(r_\perp) \left[J_{2k}(r_\perp \delta_\perp) -  \frac{2k+1}{r_\perp \delta_\perp}J_{2k+1}(r_\perp \delta_\perp)\right]\,.
\label{eq:amp}
\ee
This is the result that will be used in the numerical computations in Sec.~\ref{sec:num} where we will be keeping only the lowest $k = 0$ mode. The photo-production cross section is obtained as
\be
\frac{\rmd \sigma}{\rmd |t|} = \frac{1}{16\pi}\left|\langle \calM\rangle\right|^2\,.
\label{eq:cs}
\ee

It is instructive to provide an estimate of \eqref{eq:cs} at leading twist. In Appendix~\ref{app:oddinit} we have performed a model computation of the Odderon distribution and more details can be found in  Sec.~\ref{sec:init}. Restricting to the first non-trivial Fourier mode we find
\be
\calO_1(r_\perp,\Delta_\perp) \simeq -\frac{3\pi \rmi}{8}\frac{C_{3F}}{N_c}\alpha_S^3 r_\perp^3 A \Delta_\perp T_A(\Delta_\perp)\,,
\ee
where $T_A(\Delta_\perp)$ is the Fourier transform of the transverse profile of the target $T_A(\bp)$, see \eqref{eq:rhoA} below. $C_{3F}$ is defined in \eqref{eq:cfc3f}. Taking the limit $m_c \to \infty$ the cross section \eqref{eq:cs} is obtained in the following form
\be
\frac{\rmd \sigma}{\rmd |t|} \simeq \frac{9\pi q_c^2 \alpha \alpha_S^6 A^2  C_{3F}^2\mathcal{R}_\calP^2(0)}{4 N_c m_c^5}\frac{ |t| T_A^2(\sqrt{|t|})}{m_c^4}\,,
\label{eq:oddlt}
\ee
and so the Odderon cross section gets enhanced by $\sim A^2$ in case of nuclear targets. To get this result we have used \cite{Pham:2007xx,Dumitru:2019qec}
\be
N_c\int_z \frac{\phi_\calP(z,0)}{z\bar{z}} = \sqrt{\frac{N_c}{32\pi m_c^3}} \calR_\calP(0) \,,
\ee
where $\calR_\calP(0)$ is the radial wave function at the origin.

\subsection{The Primakoff process}

The Primakoff process corresponds to a situation with an odd number of photons instead of gluons exchanged from the target. Intuitively, we would expect the Primakoff effect to be most important in the region $\delp \simeq 0$ due to the long-range Coulomb tail of the charged target. As in the previous Sec.~\ref{sec:odder} we work in the eikonal approximation for the target interaction, with photons instead of gluons in the Wilson lines \cite{Baltz:2001dp,Gelis:2001da,Gelis:2001dh}. We thus write
\be
2\rmi \Omega(\rp,\bp) \equiv U^\dag(\xp)U(\yp) - U^\dag(\yp)U(\xp)\,,
\label{eq:photon}
\ee
in place of $\calO(\rp,\bp)$.
Here
\be
U(\xp) = \exp\left[-\frac{\rmi  e^2 q_c Z T_A(\xp)}{\boldsymbol{\pd}_\perp^2}\right] = \exp\left[4\pi \rmi q_c Z\alpha \int_{\kp}\frac{T_A(k_\perp)}{\kp^2}\rme^{\rmi \kp \cdot\xp}\right]\,,
\label{eq:wilsonphot}
\ee
is a Wilson line accounting for multiple scattering on a electromagnetic field of the target $-Ze T_A(\xp)/\boldsymbol{\pd}_\perp^2$ \cite{Baltz:2001dp,Gelis:2001da,Gelis:2001dh}. Here the transverse charge density is given as $Z T_A(\xp)$.
Because of the $\alpha$ suppression we are ignoring multiple scatterings and expand the eikonal phase to the first nontrivial order. Passing to the variable $\delp$ instead of $\bp$ we have
\be
\Omega(\rp,\delp) = -8\pi\rmi q_c Z\alpha\sin\left(\frac{\delp\cdot \rp}{2}\right) \frac{T_A(\Delta_\perp)}{\delp^2}\,,
\ee
which is the same as Eq.~(22) in \cite{Dumitru:2019qec} up to a factor due to the difference in the definition. We also obtain the Fourier moments as
\be
\Omega_{2k+1}(r_\perp,\Delta_\perp) = \frac{1}{2\pi}\int_0^{2\pi} \rmd \phi_{r\Delta} \Omega(\rp,\delp) \cos((2k+1)\phi_{r\Delta}) = -8\pi\rmi q_c Z \alpha (-1)^k J_{2k+1}\left(\frac{r_\perp \Delta_\perp}{2}\right)\frac{T_A(\Delta_\perp)}{\delp^2}\,,
\label{eq:ftphoton}
\ee
that are to be used directly in \eqref{eq:amp}.

At this point it is useful to obtain an estimate in the $m_c \to \infty$ limit, similar to what was done for the Odderon in \eqref{eq:oddlt}. We get
\be
\frac{\rmd \sigma}{\rmd |t|} \simeq \frac{\pi q_c^4 \alpha^3 Z^2 N_c \calR_\calP^2(0)}{m_c^5 |t|}T_A^2(\sqrt{|t|})\,,
\label{eq:photlt}
\ee
which displays the characteristic $1/t$ Coulomb behavior in contrast to the  Odderon case \eqref{eq:oddlt} where we have instead a suppression factor $|t|/m_c^4$. Note that $T_A(\Delta_\perp)$ is nothing but the electromagnetic charge form factor from the Rosenbluth formula \cite{Peskin:1995ev}.

In order to evaluate the Primakoff cross section numerically we must specify the
profile function $T_A(\bp)$. For the proton (neutron) targets we are replacing $Z T_A(\Delta_\perp) \to F^{p,n}_1(\Delta_\perp)$, respectively, with $F_1^{p,n}(\Delta_\perp)$ being the proton (neutron) charge form factors for which we are using a recent determination from \cite{Ye:2017gyb}.
For a nucleus we use a Woods-Saxon distribution, see \eqref{eq:rhoA} below. In this work we do not attempt to differentiate between the nuclear electromagnetic distribution and the strong interaction distribution of a nucleus, although in principle they could be different, see \cite{STAR:2022wfe,Mantysaari:2022sux}.

\section{Numerical solutions of the Odderon evolution at small-$x$}
\label{sec:bk}

Denoting the dipole distribution in the fundamental representation as
\be
\calD(\rp,\bp) \equiv \frac{1}{N_c}{\rm tr} \left\langle V^{\dag}\left(\bp + \frac{\rp}{2}\right)V\left(\bp - \frac{\rp}{2}\right) \right\rangle\,,
\ee
the fully impact parameter dependent BK equation reads \cite{Balitsky:1995ub,Kovchegov:1999yj}
\be
\frac{\pd \calD(\rp,\bp)}{\pd Y}  = \frac{\alpha_S N_c}{2\pi^2}\int_{\rop}\frac{\rp^2}{\rop^2 \rtp^2}\left[\calD(\rop,\bop)\calD(\rtp,\btp) - \calD(\rp,\bp)\right]\,,
\label{eq:bk}
\ee
where $\rtp = \rp - \rop$. In general, we have $\bop = \bp + (\rp - \rop)/2$ and $\btp  = \bp - \rop/2$ and so \eqref{eq:bk} is non-local in $\bp$. Solutions of \eqref{eq:bk} lead to unphysically large Coulomb tails in $\bp$ originating from a lack of confining interactions in the BK kernel \cite{Golec-Biernat:2003naj}. This issue has been addressed \cite{Ikeda:2004zp,Berger:2010sh,Hagiwara:2016kam,Cepila:2018faq}, at different levels of sophistication, by various modifications of the kernel in the infrared. In this work we make no attempt to tackle this difficult problem and instead resort to a local approximation $\bop \to \bp$ and $\btp \to \bp$ used in \cite{Lappi:2013zma} (see also a discussion in \cite{Kowalski:2008sa}) where the $\bp$-dependence effectively becomes an external parameter.

Splitting the dipole into Pomeron and Odderon pieces as $\calD(\rp,\bp) = 1 - \calN(\rp,\bp) + \rmi\calO(\rp,\bp)$
leads to \cite{Kovchegov:2003dm,Hatta:2005as}
\be
\begin{split}
\frac{\pd \calN(\rp,\bp)}{\pd Y} = \int_{\rop} \mathcal{K}_{\rm Bal}(\rp,\rop,\rtp)&\big[\calN(\rop,\bp) + \calN(\rtp,\bp) - \calN(\rp,\bp)\\
& + \calN(\rop,\bp)\calN(\rtp,\bp) - \calO(\rop,\bp)\calO(\rtp,\bp)\big]\,,
\label{eq:bkN}
\end{split}
\ee
\be
\begin{split}
\frac{\pd \calO(\rp,\bp)}{\pd Y} = \int_{\rop} \mathcal{K}_{\rm Bal}(\rp,\rop,\rtp)&\big[\calO(\rop,\bp) + \calO(\rtp,\bp) - \calO(\rp,\bp)\\
& - \calN(\rop,\bp)\calO(\rtp,\bp) - \calO(\rop,\bp)\calN(\rtp,\bp)\big]\,.
\label{eq:bkO}
\end{split}
\ee
In the above Eqs.~\eqref{eq:bkN}, \eqref{eq:bkO} we have replaced the conventional BK kernel with the running-coupling kernel (according to the Balitsky's prescription) \cite{Balitsky:2006wa}
\be
\frac{\alpha_S N_c}{2\pi^2}\frac{\rp^2}{\rop^2 \rtp^2} \to \mathcal{K}_{\rm Bal}(\rp,\rop,\rtp) = \frac{\alpha_S(\rp^2) N_c}{2\pi^2} \left[\frac{1}{\rop^2}\left(\frac{\alpha_S(\rop^2)}{\alpha_S(\rtp^2)} - 1\right) + \frac{\rp^2}{\rop^2 \rtp^2} + \frac{1}{\rtp^2}\left(\frac{\alpha_S(\rtp^2)}{\alpha_S(\rop^2)} - 1\right)\right]\,,
\ee
that will be used in our numerical computations. Here
\be
\alpha_S(\rp^2) = \frac{12\pi}{(33 - 2N_f)\log\left(\frac{4C^2}{\rp^2 \Lambda_{\rm QCD}^2} + \hat{a}\right)}\,,
\ee
with \cite{Lappi:2013zma} $N_f = 3$, $C^2 = 7.2$, $\Lambda_{\rm QCD} = 0.241$ GeV and $\hat{a}$ is a parameter determined by the condition $\lim_{\rp^2 \to \infty}\alpha_S(\rp^2) = \alpha_{\rm fr}$ where $\alpha_{\rm fr} = 0.7$. 

A similar system of equations was solved in \cite{Motyka:2005ep,Lappi:2016gqe,Yao:2018vcg,Braun:2020vmd,Contreras:2020lrh}, but the $b_\perp$ dependence was not addressed. Nevertheless,
some generic conclusions from these works 
also apply to our computations. Thanks to the non-linearity of the BK equation \eqref{eq:bk}, the Pomeron and the Odderon do not evolve separately. Only in the small-$r_\perp$ limit where $\calN(\rp,\bp) \to 0$ the nonlinear terms in \eqref{eq:bkO} can be neglected and the system is decoupled. When this happens, the first two terms in the square brackett \eqref{eq:bkO} cancel each other and the Odderon will become exponentially suppressed in rapidity \cite{Kovchegov:2003dm,Lappi:2016gqe,Yao:2018vcg}. In contrast, in the large $r_\perp$ region, where $\calN(\rp,\bp) \to 1$, the nonlinear terms play an important role to 
cancel the first and the second term in the square bracket in \eqref{eq:bkO} causing again an exponential suppression \cite{Kovchegov:2003dm,Motyka:2005ep,Lappi:2016gqe,Yao:2018vcg}. Such a lack of geometric scaling seems to be a general feature of not only the Odderon but also higher dipole moments in general \cite{Hagiwara:2016kam}.

\subsection{Initial conditions}
\label{sec:init}

For the Pomeron initial conditions we use a fit to HERA data from Ref.~\cite{Lappi:2013zma}. Therein, the Pomeron for the proton is modelled as
\be
\calN(\rp,\bp) = 1 - \exp\left[-\frac{1}{4}\rp^2 Q_{0,p}^2(\rp,\bp)\right]\,,
\label{eq:N0}
\ee
where
\be
Q_{0,p}^2(\rp,\bp) \equiv T_p(\bp)\frac{\sigma_0}{2} Q_{S,0}^2 \log\left(\frac{1}{r_\perp \Lambda_{\rm QCD}} + e_c \rme\right)\,,
\label{eq:Q0}
\ee
\be
T_p(\bp) = \frac{1}{\pi R_p^2} e^{-\bp^2/R_p^2}\,.
\label{eq:gaussgluon}
\ee
where we pick up $R_p$ from the relationship $\pi R_p^2 = \sigma_0/2 = 4\pi B_p$. In a recent work by Dumitru, M\"antysaari and Paatelainen \cite{Dumitru:2022ooz} the Odderon for a proton target was calculated starting from quark light-cone wavefunctions at NLO. We refer to this as the DMP model and employ it in our numerical computations.

In case of a nucleus we use again the results from Ref.~\cite{Lappi:2013zma}, with the Pomeron distribution given as in \eqref{eq:N0} but
with
\be
Q_{0,A}^2(\rp,\bp) \equiv A T_A(\bp)\frac{\sigma_0}{2} Q_{S,0}^2 \log\left(\frac{1}{r_\perp \Lambda_{\rm QCD}} + e_c \rme\right)\,,
\label{eq:Q0A}
\ee
in place of $Q_{0,p}^2(\rp,\bp)$. $T_A(\bp)$ is the transverse profile of a nuclear target. The parameters in \eqref{eq:Q0} are given as $Q_{S,0}^2 = 0.06$ GeV$^2$,  $e_c = 18.9$ and $\frac{\sigma_0}{2} = 16.36$ mb \cite{Lappi:2013zma}. $T_A(\bp)$ is obtained by integrating the Woods-Saxon distribution \cite{Lappi:2013zma}
\be
T_A(\bp) = \int_{-\infty}^\infty \rmd z \frac{n_A}{1 + \exp\left[\frac{\sqrt{\bp^2 + z^2} - R_A}{d}\right]}\,,
\label{eq:rhoA}
\ee
which is normalized to unity $\int_{\bp} T_A(\bp) = 1$. This fixes $n_A$ as $-8\pi n_A d {\rm Li}_3(-\rme^{R_A/d}) = 1$. 
Here $d = 0.54$ fm, $R_A = 1.12 A^{1/3} - 0.86 A^{-1/3}$ fm \cite{Lappi:2013zma}. These Woods-Saxon parameters are numerically very close to the fit values from \cite{DeVries:1987atn}.

The initial condition of the Odderon for a nuclear target is based on the Jeon-Venugopalan (JV) model \cite{Jeon:2005cf}, which involves a cubic term added to the standard McLerran-Venugopalan small-$x$ functional
\be
W[\rho] = \exp\left[-\int_{\xp} \left(\frac{\delta_{ab}\rho^a(\xp)\rho^b(\xp)}{2\mu^2} - \frac{d_{abc}\rho^a(\xp)\rho^b(\xp)\rho^c(\xp)}{\kappa}\right)\right]\,,
\label{eq:jv}
\ee
where
\be
\mu^2 = \frac{g^2}{2}\frac{A}{\pi R_A^2} \,, \qquad \kappa = g^3 N_c\frac{A^2}{(\pi R_A^2)^2}\,.
\label{eq:couplings}
\ee 
In \cite{Jeon:2005cf} (see also \cite{Kovchegov:2003dm}), it was found that the Odderon distribution from the above functional takes the following form
\be
\calO(\xp,\yp) = - g^3 C_{3F} \frac{\mu^6}{\kappa}\Theta(\xp,\yp)\exp\left[-\frac{g^2 C_F \mu^2}{2}\Gamma(\xp,\yp)\right]\,,
\label{eq:oddjv}
\ee
where 
\be
C_F = \frac{N_c^2 - 1}{2N_c}\,, \qquad C_{3F} = \frac{(N_c^2 - 1)(N_c^2 - 4)}{4N_c^2}\,,
\label{eq:cfc3f}
\ee
and
\be
\begin{split}
&\Gamma(\xp,\yp) = (\pi R_A^2)\int_{\zp}T_A(\zp) \left[G(\xp - \zp) - G(\yp - \zp)\right]^2\,,\\
&\Theta(\xp,\yp) = (\pi R_A^2)\int_{\zp} T_A(\zp) \left[G(\xp - \zp) - G(\yp - \zp)\right]^3\,,
\end{split}
\label{eq:gth}
\ee
where $G(\xp-\zp)$ is a 2D Green function \eqref{eq:2dgf} and we have inserted the target profile $T_A(\bp)$,
see the discussion in the Appendix \ref{app:oddinit}. Eq.~\eqref{eq:oddjv} can be interpreted as a single perturbative Odderon with any number of perturbative Pomeron insertions. Starting from \eqref{eq:oddjv} we deduce the following result for the Odderon initial condition
\be
\begin{split}
\calO(\rp,\bp) &= \frac{\lambda}{8} \left[R_A\frac{\rmd T_A(\bp)}{\rmd b_\perp} A^{2/3}\frac{\sigma_0}{2}\right] Q^3_{S,0}A^{1/2} r_\perp^3 (\hat{\boldsymbol{r}}_\perp\cdot\hat{\boldsymbol{b}}_\perp)\log\left(\frac{1}{r_\perp \Lambda_{\rm QCD}} + e_c \rme\right) \exp\left[-\frac{1}{4}\rp^2 Q_{0,A}^2(\rp,\bp)\right]\,,
\end{split}
\label{eq:oddinit}
\ee
where in the JV model we would have
\be
\lambda_{\rm JV} = -\frac{3}{16}\frac{N_c^2 - 4}{(N_c^2 - 1)^2}\frac{Q_{S,0}^3 A^{1/2} R_A^3}{\alpha_S^3 A^2}\,.
\label{eq:lamjv}
\ee
The details of the computation leading to \eqref{eq:oddinit} are given in the Appendix \ref{app:oddinit}.

\subsection{Numerical solutions}

\begin{figure}
  \begin{center}
  \includegraphics[scale = 0.4]{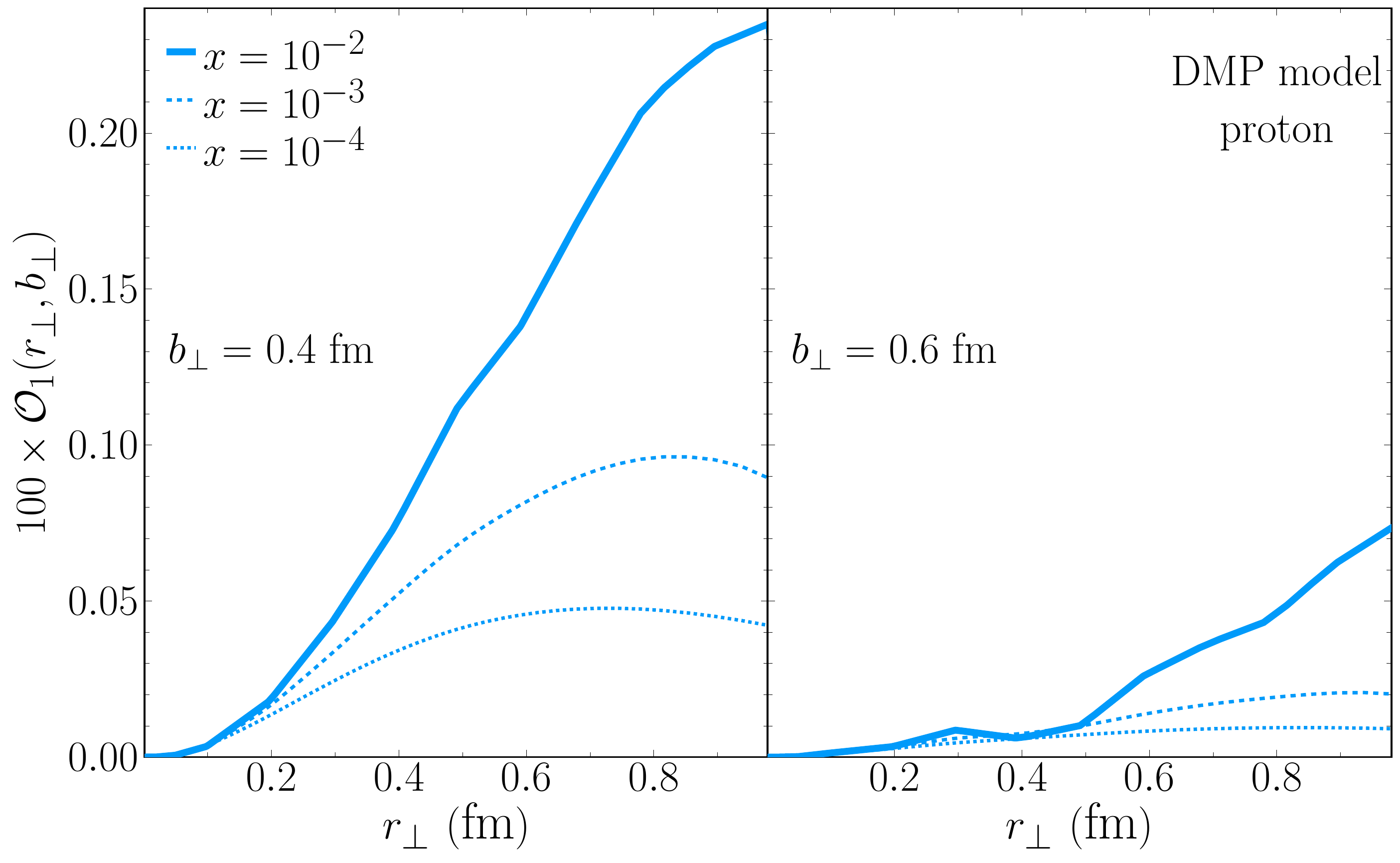}
  \end{center}
  \caption{The first Fourier moment $O_1(r_\perp,b_\perp)$ of the Odderon distribution of the proton in the DMP model as a function of $r_\perp$ for different values of $x$ and at the impact parameters $b_\perp = 0.6$ fm and $0.4$ fm.}
  \label{fig:Odderonr}
\end{figure}

\begin{figure}
  \begin{center}
  \includegraphics[scale = 0.4]{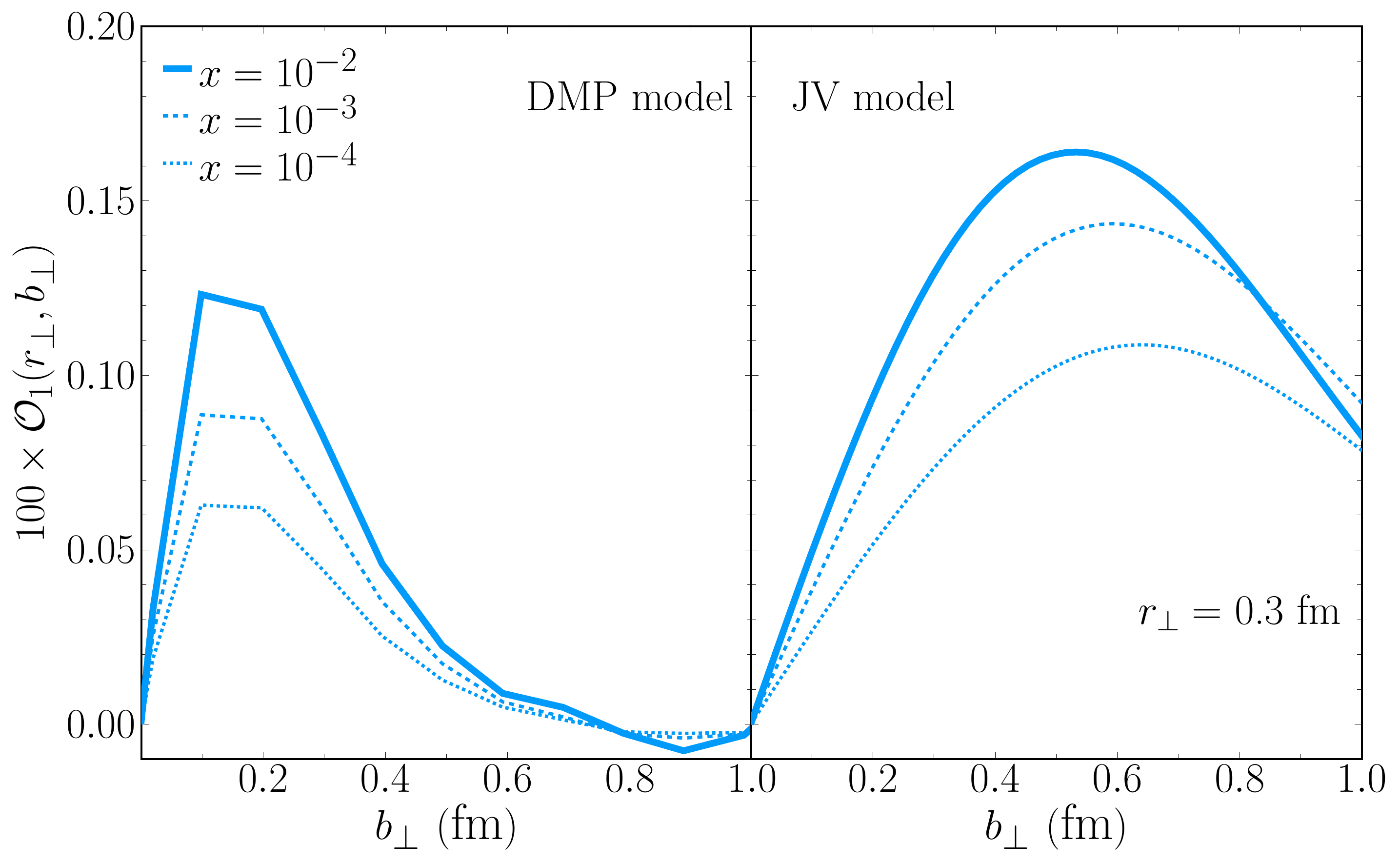}
  \end{center}
  \caption{Left: the first Fourier moment $O_1(r_\perp,b_\perp)$ of the Odderon distribution of the proton in the DMP model as a function of $b_\perp$ for different values of $x$. Right: same quantity, but in the JV model.}
  \label{fig:Odderonb}
\end{figure}

The system of BK equations \eqref{eq:bkN}-\eqref{eq:bkO} was solved on a $(r_\perp,b_\perp,\phi_{rb})$ grid, where $\phi_{rb} = \phi_r - \phi_b$. As mentioned earlier, we consider $b_\perp$ as an external parameter and solve the BK equation for each value of $b_\perp$ separately. The integral over $\rop$ in the equations \eqref{eq:bkN} and \eqref{eq:bkO} is evaluated over a lattice in $(r_\perp, \phi_{rb})$  using adaptive cubature \cite{GENZ1980, GENZ1991}. The lattice is equally spaced in $\log r_\perp$ from $r_\perp=10^{-6}\text{ GeV}^{-1}$ to $10^{4}\text{ GeV}^{-1}$ with $n_{r_\perp}=500$ lattice points and in $\phi_{rb}$ from $\phi_{rb}=0$ to $2\pi$ with $n_{\phi_{rb}}=100$ lattice points.  For each value of $b_\perp$, the equations \eqref{eq:bkN} and \eqref{eq:bkO} together represent a system of $2\times n_{r_\perp}\times n_{\phi_{rb}}$ coupled differential equations representing the values of the Pomeron and the Odderon over the grid. This system of differential equations is solved using a three-step third order Adams-Bashforth method with a step size in rapidity $\Delta Y=0.1$ for up to $Y=5$. The first two timesteps required to initiate the Adams-Bashforth method were obtained using Ralston's second order method. We have validated our numerical treatment of the BK system in two ways. First, since we have adopted our parametrization of the Pomeron from \cite{Lappi:2013zma}, we have checked that our results for the BK evolved the dipole amplitude in the proton and in the nuclei agree with \cite{Lappi:2013zma}.
Second, we checked that we were able to reproduce fully the results for the BK evolution of the spin-dependent Odderon presented in \cite{Yao:2018vcg}. We additionally checked several different methods for solving the BK system (including the Euler method, a range of Adam-Bashforth methods, and the fourth order Runge-Kutta method) and found the third-order Adams-Bashforth method to be optimal.

At this point we make a comment about the angular dependence.
The Pomeron initial condition \eqref{eq:N0} is independent of $\phi_{rb}$, while the $\cos(\phi_{rb})$ moment in the Odderon initial condition \eqref{eq:oddinit} will generate a $\cos(2\phi_{rb})$ moment in the Pomeron through the $\sim \calO^2$ term in \eqref{eq:bkN}. In principle, this further backreacts onto the Odderon through the $\sim\calN \calO$ pieces generating a higher $\cos(3\phi_{rb})$ moment in the Odderon. However, in our numerical computation we find that already the $\cos(2\phi_{rb})$ term is numerically tiny in support of the similar findings reported in \cite{Motyka:2005ep,Yao:2018vcg}\footnote{In particular, this also implies the HERA fit \cite{Lappi:2013zma} of the Pomeron does not get affected by the presence of the Odderon in the BK equation.}. For this reason, in the following results we will discuss the Odderon solution only in the context of its dominant $\calO_1(r_\perp,b_\perp)$ moment.

On Fig.~\ref{fig:Odderonr} we show the first Odderon moment $\calO_1(r_\perp,b_\perp)$ for the proton target using the DMP model as initial condition as a function of $r_\perp$ for several finite values of $b_\perp$. Going from the full line at the initial condition $x = 10^{-2}$ the Odderon is severely affected in magnitude when evolving to smaller $x$ as can be seen by the thin dashed curve where $x = 10^{-3}$ and a thin dotted curve where $x = 10^{-4}$, verifying numerically the lack of geometric scaling for the Odderon. Moving on to the $b_\perp$ dependence, the left plot on Fig.~\ref{fig:Odderonb} shows $\calO_1(r_\perp,b_\perp)$ as a function $b_\perp$ with $r_\perp$ fixed and for different values of $x$. For illustrative purposes we plot on the right the result for the proton target as obtained in the JV model. Interestingly, while the DMP model Odderon is peaked within the proton the JV model Odderon is peaked at higher $b_\perp$ due to the $\sim d T_p /db_\perp$ term.

\begin{figure}
  \begin{center}
  \includegraphics[scale = 0.45]{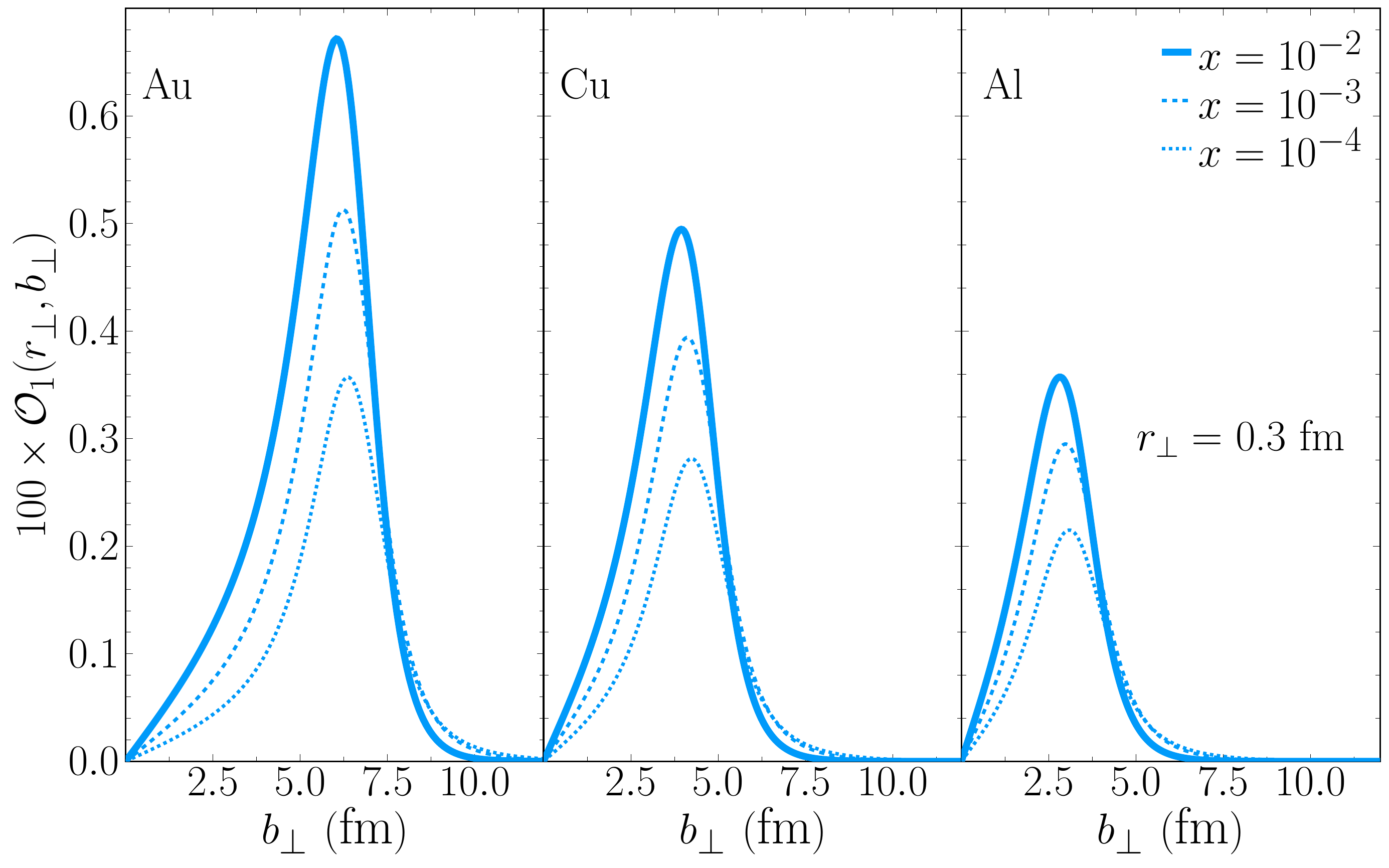}
  \end{center}
  \caption{The first Fourier moment $O_1(r_\perp,b_\perp)$ of the Odderon distribution of the nuclei in the JV model as a function of $b_\perp$ for different values of $x$. Left plot is for the Au, center is for Cu and right is for Al nuclei.}
  \label{fig:Odderonnucl}
\end{figure}

Comparing the results in the DMP and the JV models, we can quantify some of the model uncertainties concerning the magnitude of the Odderon. For this purpose we take the absolute ratio of the $\eta_c$ production amplitudes in the DMP and the JV models in the case of the nucleon target. In the limit $\Delta_\perp \to 0$, and for $Q^2 = 1$ GeV$^2$, we find $\langle \calM\rangle_{p, \rm \,DMP} /\langle \calM\rangle_{p, \rm \, JV} \to 0.026$.
On the other hand, an upper bound on the Odderon is imposed by the group theory constraint \cite{Kaiser_2006,Lappi:2016gqe}
\be
\left(4-3\calN(\rp,\bp)\right)\calN^3(\rp,\bp) - 6\left(6 - 6\calN(\rp,\bp) + \calN^2(\rp,\bp)\right)\calO^2(\rp,\bp) - 3 \calO^4(\rp,\bp) \geq 0\,.
\label{eq:bound}
\ee
In the small-$r_\perp$ limit this simplifies to $\calO^2(\rp,\bp) \leq \calN^3(\rp,\bp)/9$ \cite{Lappi:2016gqe}. We have checked that the DMP model satisfies this bound. Using the JV initial condition  for nuclei we can quantify \eqref{eq:bound} as a bound on the magnitude of $\lambda$ and numerically we find that model coupling is somewhat below the bound, namely
\be
\lambda_{\rm max}^{197} = 1.143\lambda_{\rm JV}^{197}\,,\qquad \lambda_{\rm max}^{63} = 1.553\lambda_{\rm JV}^{63}\,, \qquad \lambda_{\rm max}^{27} = 2.26\lambda_{\rm JV}^{27}\,.
\label{eq:boundnum}
\ee
where $\lambda_{\rm JV}$ is given by \eqref{eq:lamjv} and the superscript refers to the atomic number for different species of nuclei. We have checked that \eqref{eq:bound} is satisfied for all $r_\perp$ and $b_\perp$, where for the latter, we considered the domain for which the nuclear saturation scale is above the minimum bias saturation scale of the proton. We will thus consider $\lambda$ up to $\lambda_{\rm max}$. For orientation purposes, the lowest coupling we consider for nuclei will be given as $\lambda = 0.026\lambda_{\rm JV}$, where the proportionality factor $0.026$ is fixed by the DMP vs. JV amplitude ratio for the proton target discussed above.

Finally, on Fig.~\ref{fig:Odderonnucl} we show the results for the $b_\perp$ dependence of the $\calO_1(r_\perp,b_\perp)$ for the nuclear targets: Au (left), Cu (center) and Al (right) using the JV model. Evolving to smaller values in $x$, the peak in the Odderon distribution drops in magnitude but also shifts to slightly larger $b_\perp$. This will leave an interesting consequence in the diffractive pattern of the cross section as we will explain in the following Section \ref{sec:num}.

\section{Numerical results for the cross section}
\label{sec:num}

\begin{figure}
  \begin{center}
  \includegraphics[scale = 0.4]{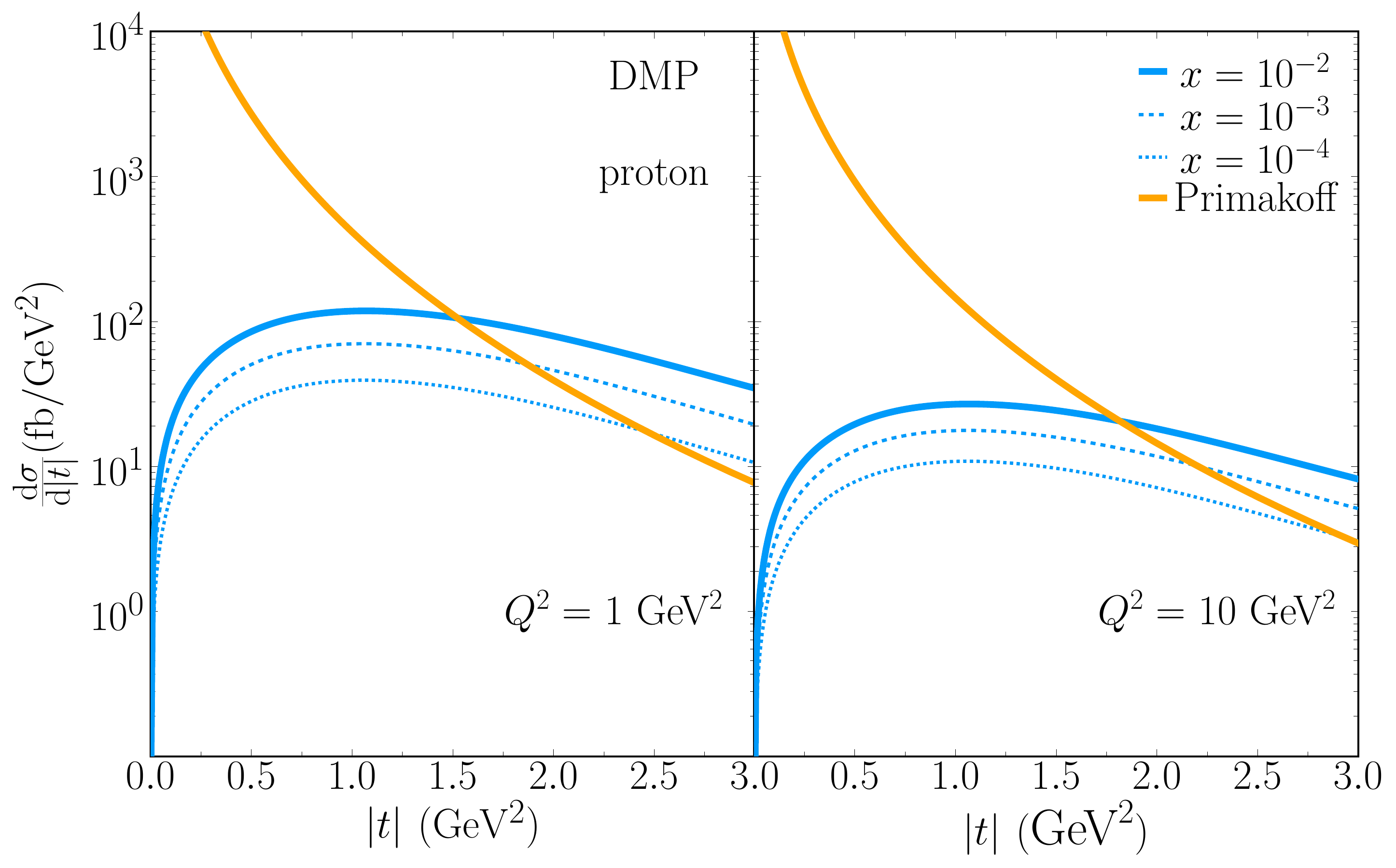}
  \end{center}
  \caption{$|t|$ dependence of the $\gamma^* p \to \eta_c p$ cross section with the DMP model. The contribution from the Primakoff process is shown separately.}
  \label{fig:protont}
\end{figure}

\begin{figure}
  \begin{center}
  \includegraphics[scale = 0.4]{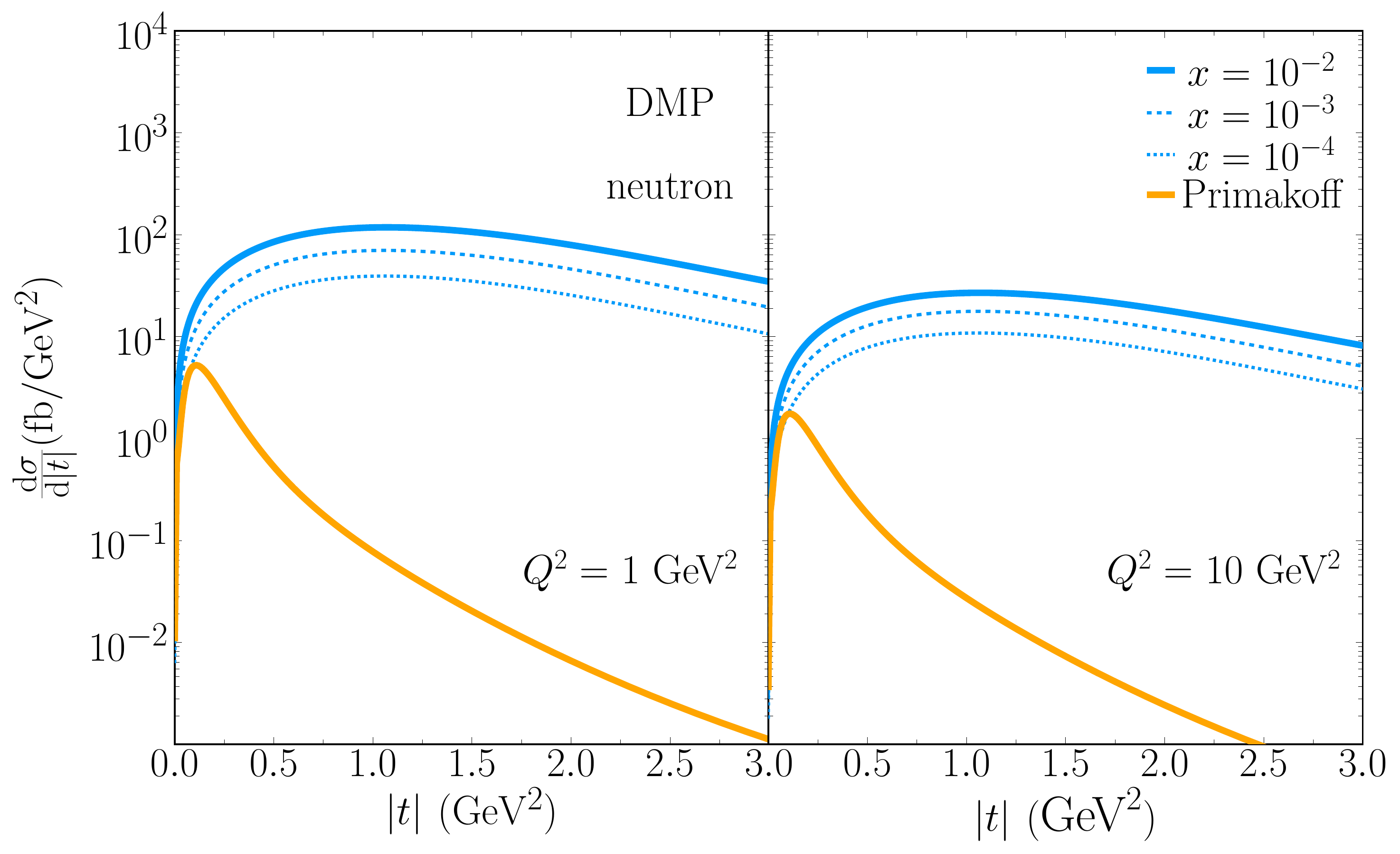}
  \end{center}
  \caption{$|t|$ dependence of the $\gamma^* n \to \eta_c n$ cross section with the DMP model. The contribution from the Primakoff process is shown separately.}
  \label{fig:neutront}
\end{figure}

In this Section we show the results of the numerical computation of the photoproduction cross section for the exclusive processes $\gamma^* p \to \eta_c p$, $\gamma^* n \to \eta_c n$ and $\gamma^* A \to \eta_c A$, where we consider the Au, Cu and Al nuclei.
The numerical computation of the cross section \eqref{eq:cs} is based on the amplitude for the Odderon contribution given by \eqref{eq:amp}. To compute the Primakoff cross section we use the same Eq.~\eqref{eq:amp} with the replacement $\calO_{2k+1}(r_\perp,\Delta_\perp) \to \Omega_{2k+1}(r_\perp,\Delta_\perp)$, where $\Omega_{2k+1}(r_\perp,\Delta_\perp)$ is given by \eqref{eq:ftphoton}. In all the computations considered, we restrict to the lowest $k = 0$ Fourier moment of the amplitude. We have explicitly checked that the contributions from the higher moments are strongly suppressed both in the case of the Odderon and the Primakoff contributions relative to the $k = 0$ case. For the Fourier transform in the impact parameter $\bp$ we used the Ogata quadrature method \cite{Kang:2019ctl}.

We first discuss the numerical results for exclusive $\gamma^*p \to \eta_c p$ photoproduction. Fig.~\ref{fig:protont} shows the cross-section as a function of $|t|$ for several values of $x$ and $Q^2$. The computation is performed using the DMP model. The result shows a rather small $|t|$-slope of the cross section. This is a generic feature of the quark based approach as the three gluons in the Odderon can couple to three different quarks leaving the proton intact even at relatively large momentum transfer \cite{Czyzewski:1996bv,Dumitru:2020fdh}.  The Primakoff cross section overwhelms the Odderon cross section at small $|t|$, but this gets reversed for $|t| \gtrsim$ 1.5 GeV$^2$ thanks to a small $|t|$-slope of the Odderon cross section.

The small-$x$ evolution reduces the Odderon cross section by roughly an order of magnitude when going from $x \sim 10^{-2}$ to $x\sim 10^{-4}$. However, it is still above the Primakoff background for $|t| \gtrsim $ 2-3 GeV$^2$, with the $|t|$-slope remaining roughly the same. Our conclusion for proton targets is thus similar to that of \cite{Dumitru:2019qec} where the computation was performed at moderate $x \sim 0.1$. The Odderon extraction from collisions on the proton target would thus require measurements of the cross section at potentially large momentum transfers even when $x$ is small $x \lesssim 0.01$.  For neutron targets the Primakoff cross section is only a very small contribution and the Odderon can be probed even at low $|t|$ and/or low $x$ -- see Fig.~\ref{fig:neutront}.

\begin{figure}
  \begin{center}
  \includegraphics[scale = 0.5]{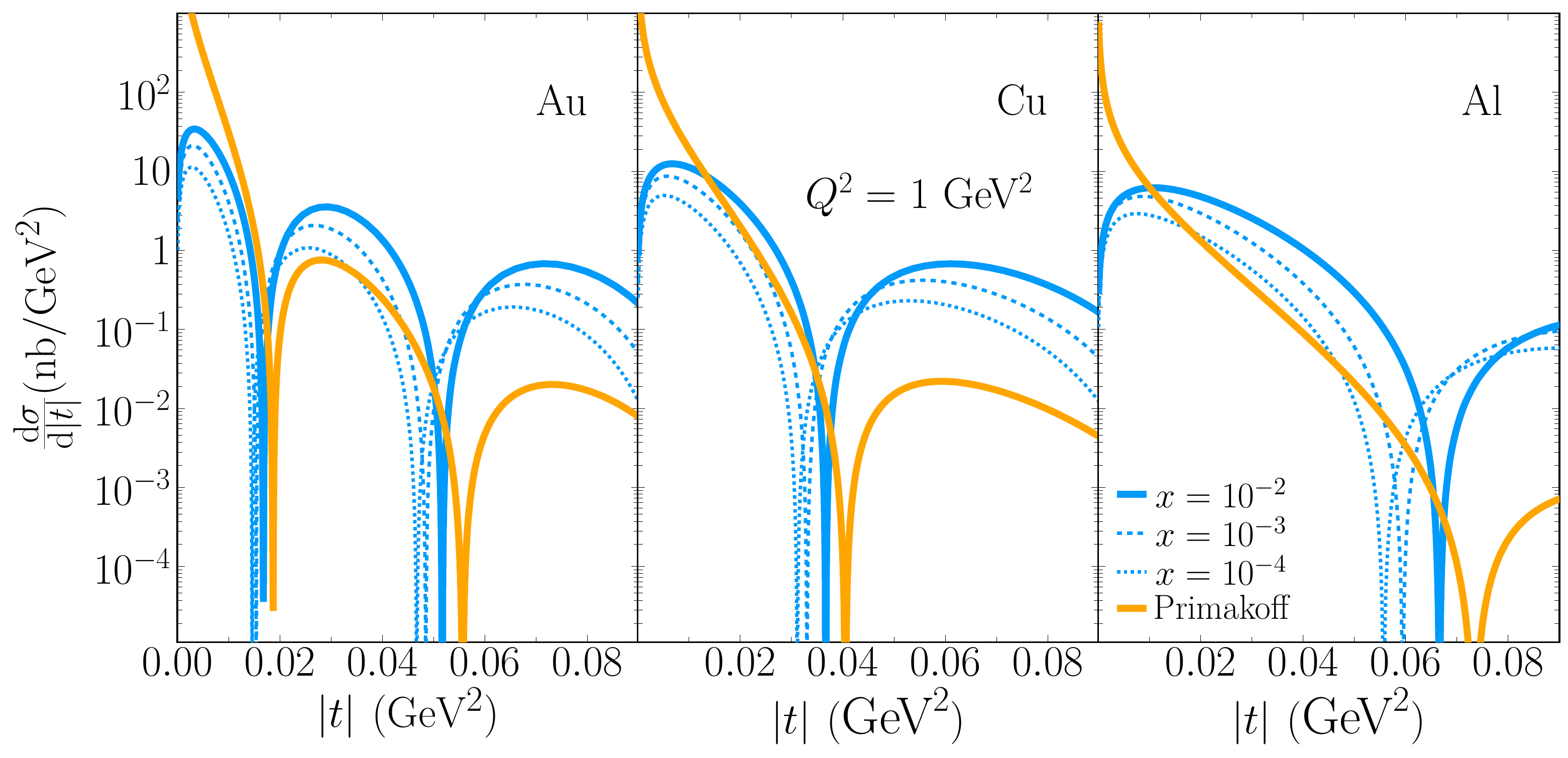}
  \end{center}
  \caption{The $\gamma^* A \to \eta_c A$ cross section for three different targets: Au (left), Cu (center) and Al (right). The odderon coupling is fixed to the maximal value allowed by the group theory constraint \eqref{eq:boundnum}.}
  \label{fig:nucleit}
\end{figure}

On Fig.~\ref{fig:nucleit} we show the numerical results for the $\gamma^* A \to \eta_c A$ cross section for Au (left), Cu (center) and Al (right) targets. The Odderon coupling $\lambda$ is set to the maximal value allowed by the group theory constraint \eqref{eq:boundnum}. The Odderon (and the Primakoff) cross section become enhanced by the mass (atomic) number of the target. For example, using maximal coupling allowed by the group theory constraint ($\lambda = \lambda_{\rm max}$), the Odderon cross section can reach up to about 10 nb/GeV$^2$ for Au. Taking instead $\lambda = 0.026 \lambda_{\rm JV}$ (the factor $0.026$ is determined by the DMP vs JV amplitude ratio) as an assumption for the lowest estimate, leads to $\sim 5$ pb/GeV$^2$.

Both the Odderon and the Primakoff contributions show characteristic diffractive patterns that are mostly of a geometric origin. However, it is clearly visible that the diffractive pattern for the Odderon cross section is altered compared to the Primakoff case: the diffractive dips are shifted to smaller $|t|$ even for the initial condition and the shift becomes more pronounced as $x$ gets smaller or $|t|$ gets larger. To understand this result, notice that according to the leading twist estimates in \eqref{eq:oddlt} and \eqref{eq:photlt} the Odderon and the Primakoff cross sections behave as $\rmd \sigma /\rmd|t| \propto |t| T^2_A(\sqrt{|t|})$ and  $\rmd \sigma /\rmd|t| \propto T^2_A(\sqrt{|t|})/|t|$, respectively.
We are lead to the conclusion that the shift of the diffractive pattern when comparing the Odderon and the Primakoff cross section is a consequence of multiple scatterings in the Odderon amplitude. This finds additional support by the evolution to smaller $x$ where, as a consequence of the growth of the saturation scale, multiple scattering effects become increasingly important, acting to further increase the shift.

\begin{figure}
  \begin{center}
  \includegraphics[scale = 0.5]{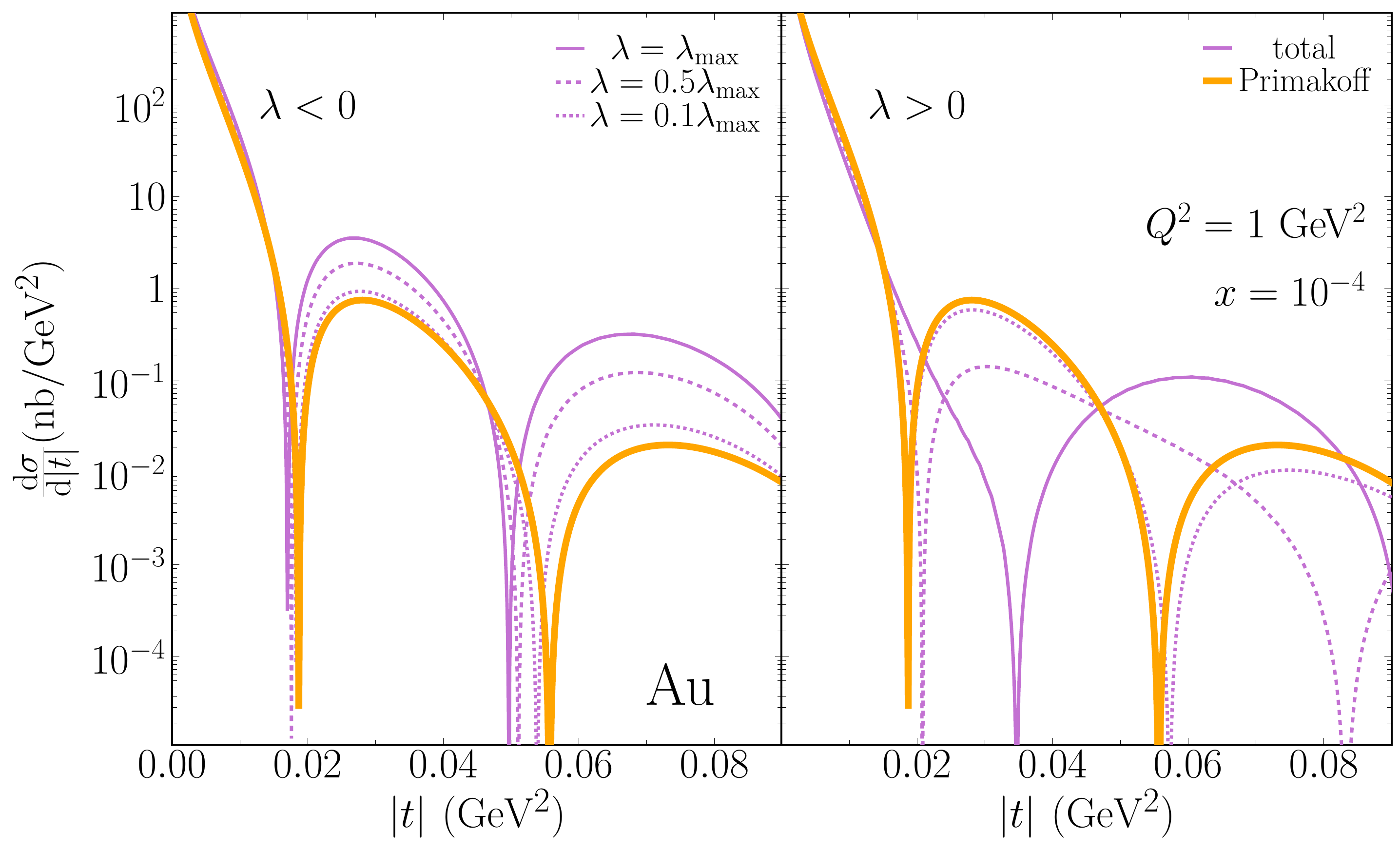}
  \end{center}
  \caption{The $\gamma^* {\rm Au} \to \eta_c {\rm Au}$ cross section for three considered values of the odderon coupling up to the maximal value allowed by the group theory constraint \eqref{eq:bound}. On the left (right) panel the sign of the Odderon coupling parameter is chosen as $\lambda < 0$ ($\lambda > 0$). The purple curves stand for the total cross section, with individual line styles representing different values of $\lambda$.}
\label{fig:nuclei2}
\end{figure}

Considering the total cross section, where the Odderon and the Primakoff contributions must be added coherently, the relative sign between the two amplitudes determines whether they interfere constructively or destructively. In our computation this is controlled by the sign of the Odderon coupling parameter $\lambda$. Using the JV model the sign is negative, see \eqref{eq:lamjv}. Thanks to the $\rmd T_A/\rmd b_\perp$ term, this gives a positive $\calO_1(r_\perp,b_\perp)$ overall, see Fig.~\ref{fig:Odderonnucl}. For comparison, the DMP model computation for proton targets \cite{Dumitru:2022ooz} also yields a positive $\calO_1(r_\perp,b_\perp)$, see Fig.~\ref{fig:Odderonb}. While positive $\calO_1(r_\perp,b_\perp)$ seems to be preferred by model computations, on Fig.~\ref{fig:nuclei2} we compute the total cross section considering both signs of $\calO_1(r_\perp,b_\perp)$ (or, equivalently, $\lambda$). For $\calO_1(r_\perp,b_\perp)>0$ ($\lambda < 0$) the results are given on the left panel of Fig.~\ref{fig:nuclei2}. In this case the interference of the Odderon and Primakoff amplitudes is mostly constructive. Our result demonstrates that the multiple scattering effect in the Odderon amplitude, that shifts the diffractive pattern relative to the Primakoff component, can leave its trace also in the total cross section depending on the magnitude of the Odderon. On the right panel of Fig.~\ref{fig:nuclei2}, the opposite case of $\calO_1(r_\perp,b_\perp)<0$ ($\lambda > 0$) is displayed. The two amplitudes are now out of phase and interfere destructively, resulting in a severe distortion of the diffractive pattern in the total cross section in comparison to the Primakoff contribution only. We conclude that in both cases the known Primakoff diffractive dips could be filled in the total cross section. This could be used as a signal of the Odderon from exclusive $\eta_c$ production off nuclear targets. Considering different nuclear species could be a valuable tool in verifying this suggestion.

\section{Conclusion}
\label{sec:conc}

In this work we have computed the exclusive $\eta_c$ production in $ep$ and $eA$ collisions as a potential probe of the Odderon. Our computation relies on the CGC formalism where the effect of multiple scatterings is taken explicitly into account in a description of scattering off a dense target at small-$x$. We have numerically solved the BK evolution equation in impact parameter $b_\perp$ and dipole size $r_\perp$ for the coupled Pomeron-Odderon system. The numerical results demonstrate a rapid drop in the Odderon with evolution in line with the results in the literature \cite{Kovchegov:2003dm,Lappi:2016gqe,Yao:2018vcg}. 

Due to a large Primakoff background we find that in order to isolate the Odderon component of the cross section for the proton target, it is required to have relatively large momentum transfers: $|t| \gtrsim$ 1.5-3 GeV$^2$ for $x \sim 10^{-2}-10^{-4}$. On a qualitative level this is rather similar to the conclusions drawn in the previous works \cite{Engel:1997cga,Czyzewski:1996bv,Bartels:2001hw,Ma:2003py,Dumitru:2019qec}. A new result is that the $|t|$-slope is not altered by small-$x$ evolution, although the cross section does reduce in magnitude. Exclusive scattering off a neutron leads to a negligible Primakoff component and represents a new opportunity to probe the Odderon at low $|t|$. In practice this could be done using deuteron or He$^3$ targets with spectator proton tagging in the near forward direction, see for example \cite{CLAS:2011qvj,Friscic:2021oti}.
		
For the nuclear targets we have found that the saturation effects in the Odderon distribution distorts the diffractive pattern in comparison to the Primakoff process. The effect is a few percent in magnitude and accumulates for smaller $x$ and/or larger momentum transfers. Depending on the coupling of the Odderon, it is possible that the diffractive dips of the Primakoff process get filled by the Odderon component of the cross section. Such a distortion of the diffractive pattern in comparison to the known nuclear charge form factors might be a new way to measure the Odderon component in the nuclear wave function.


As our final remark, we wish to clearly state that the actual experimental measurement of the Odderon component of the exclusive $\eta_c$ cross section is certainly challenging. Firstly, the Odderon itself is small, and so the cross section with proton (or neutron) targets tends to be low ($\sim 10^2$ fb/GeV$^2$). This could be circumvented by considering nuclear targets instead as the Odderon cross section is enhanced roughly as $\sim A^2$. With the maximal Odderon coupling allowed by the group theory constraint the cross section can be in the range of nb/GeV$^2$. However, experimental extractions of a shift in the diffractive pattern in $\gamma^* A \to \eta_c A$, found at moderate/high $|t|$, calls for a good control of the incoherent background - a related discussion, albeit for the Pomeron, can be found in \cite{Klein:2019qfb,AbdulKhalek:2021gbh}. Secondly, the branching ratio for $\eta_c$ to charged hadrons is only a few percent \cite{Workman:2022ynf} with a serious background from feed-down of $J/\psi$ subsequently decaying as $J/\psi \to \eta_c \gamma$ with $\gamma$ undetected \cite{Czyzewski:1996bv,Klein:2018ypk,Harland-Lang:2018ytk}. Nevertheless, $\eta_c$ has been measured through its hadronic channel in $e^+ e^-$ by BABAR \cite{BaBar:2010siw} and so such difficulties might be overcome also at EIC. Measuring at least the Primakoff component seems to be a feasible starting point \cite{Babiarz:2023cac}. In any case, we consider the conclusions drawn from our results to be rather generic that would also be present in case of other quarkonia states or light mesons. 

\acknowledgments

S.~B., A.~K. and E.~A.~V. are supported by the Croatian Science Foundation (HRZZ) no. 5332 (UIP-2019-04). S.~B. thanks Adrian Dumitru and Leszek Motyka for stimulating discussions. A.~K would like to thank Brookhaven National Lab, where part of this work was performed, for their warm hospitality.

\appendix

\section{Initial condition for the Odderon}
\label{app:oddinit}

In Ref.~\cite{Jeon:2005cf} (see also \cite{Dumitru:2011zz}) Jeon and Venugopalan used a model functional with quadratic and cubic interaction \eqref{eq:jv} in order to find the Odderon operator. The parameters $\mu^2$ and $\kappa$ \eqref{eq:couplings} were treated as constants. In order to include the impact parameter dependence we assume that $\mu^2$ and $\kappa$ have a transverse profile $T_A(\xp)$ with $\int_{\xp} T_A(\xp) = 1$ such that the average couplings are given by \eqref{eq:couplings}. In this case we are lead to a straightforward generalization of (34) in \cite{Jeon:2005cf} given by \eqref{eq:oddjv}. We note in passing that in the Gaussian approximation the Pomeron $\calN(\xp,\yp)$ takes the form	
\be	
\calN(\xp,\yp) = 1-\exp\left[-\frac{g^2 C_F \mu^2}{2}\Gamma(\xp,\yp)\right]\,.	
\label{eq:pom}	
\ee	
Inserting \eqref{eq:couplings} into \eqref{eq:oddjv}, with $T_A(\zp) \to \frac{1}{\pi R_A^2}$, formally recovers (34) in \cite{Jeon:2005cf}. The 2D Green function $G(\xp - \yp)$ in \eqref{eq:oddjv} and \eqref{eq:gth} is explicitly given as	
\be	
G(\xp - \yp) = \int_{\kp} \frac{\rme^{-\rmi \kp \cdot(\xp - \yp)}}{\kp^2 + m^2}\,,	
\label{eq:2dgf}	
\ee	
with $m$ an IR cutoff. Eq.~\eqref{eq:oddjv} is the basis point to derive the initial condition for the Odderon. Its derivation essentially rests on the assumption that the cubic ($\rho^3$) term in \eqref{eq:jv} is parametrically suppressed as $A^{-1/6}$ \cite{Jeon:2005cf,Dumitru:2018vpr} for a large nuclei, as compared to the quadratic ($\rho^2$) term \cite{Jeon:2005cf} and so \eqref{eq:oddjv} is obtained by expanding to first order in $\rho^3/\kappa$ while summing in $\rho^2/\mu^2$ to all orders. 	
In the following we compute $\Gamma(\xp,\yp)$ and $\Theta(\xp,\yp)$.	
Going to momentum space we have	
\be
\begin{split}
\Gamma(\xp,\yp) &= (\pi R_A^2)\int_{\pp \kp} T_A(\pp) \rme^{-\rmi \pp \cdot \bp}\frac{1}{\kp^2 + m^2} \frac{1}{(\kp - \pp)^2 + m^2}\\
&\times\left[\rme^{-\rmi \kp \frac{\rp}{2}}- \rme^{\rmi \kp \frac{\rp}{2}}\right]\left[\rme^{-\rmi (\pp - \kp) \frac{\rp}{2}}- \rme^{\rmi (\pp - \kp) \frac{\rp}{2}}\right]\,,	
\end{split}
\ee	
where we used $\rp = \xp - \yp$ and $\bp = (\xp + \yp)/2$. Assuming $p_\perp$ is small ($b_\perp$ is large) and expanding the phase around small $\rp$ we have	
\be	
\begin{split}	
\Gamma(\xp,\yp) &\simeq (\pi R_A^2) T_A(\bp)\int_{\kp}\frac{(\kp\cdot\rp)^2}{(\kp^2 + m^2)^2} \simeq \frac{T(\bp)}{4\pi}\rp^2 \int_0^{\Lambda}  k_\perp \rmd k_\perp \frac{\kp^2}{(\kp^2 + m^2)^2}\\	
& \simeq (\pi R_A^2)\frac{T_A(\bp)}{4\pi}\rp^2 \log\left(\frac{1}{r_\perp m} + \rme\right)\,,	
\end{split}	
\label{eq:gamma}	
\ee	
where in the last equality we extracted the leading log and the UV cuttoff is placed on the $k_\perp$ integral as $\Lambda \propto 1/r_\perp$. Using \eqref{eq:gamma} in the argument of the exponential in \eqref{eq:pom} the result coincides with \cite{Lappi:2013zma} with the conventional definition	
\be	
Q_S^2 \equiv \frac{C_F g^2 \mu^2}{2\pi}\,.	
\ee

For $\Theta(\xp,\yp)$ we similarly have
\be
\begin{split}
\Theta(\xp,\yp) &\simeq (\pi R_A^2)\rmi \int_{\pp \kp \kp'} T(\pp) \rme^{-\rmi \pp \cdot\bp} (\kp \cdot \rp)(\kp' \cdot \rp)(\pp - \kp - \kp' )\cdot \rp \frac{1}{\kp^2 + m^2}\\
&\frac{1}{\kp'^2 + m^2} \frac{1}{(\pp - \kp - \kp')^2 + m^2}\,,
\end{split}
\ee
where we already expanded for $\rp \to 0$. Assuming also small $\pp$ we have
\be
\frac{(\pp - \kp - \kp')\cdot \rp}{(\pp - \kp - \kp')^2 + m^2} \simeq \frac{(\pp \cdot \rp)}{(\kp + \kp')^2 + m^2} - \frac{2\left((\kp + \kp') \cdot \rp\right) \left((\kp + \kp')\cdot \pp\right)}{\left[(\kp + \kp')^2 + m^2\right]^2}\,.
\ee
The zeroth order term above vanishes by rotation invariance. Using the second term we perform the angular integrals
\be
\begin{split}
&\int_0^{2\pi}\frac{\rmd\phi}{2\pi}\int_0^{2\pi}\frac{\rmd \phi'}{2\pi} \frac{(\kp \cdot \rp)(\kp' \cdot \rp)(\pp - \kp - \kp')\cdot \rp}{(\pp - \kp - \kp')^2 + m^2}\\
&\simeq -\frac{3}{2}(\pp \cdot \rp)\rp^2 \frac{\kp^2 \kp'^2 m^2}{\left[(\kp^2 + \kp'^2 + m^2)^2 - 4\kp^2 \kp'^2\right]^{3/2}} \equiv (\pp \cdot \rp) \rp^2 \calJ(k_\perp,k'_\perp)\,.
\end{split}
\ee
Integrating further over $k'_\perp$ leads to
\be
\begin{split}
\frac{1}{2\pi}\int_0^\infty &\frac{ \calJ(k_\perp,k'_\perp) k'_\perp\rmd k'_\perp}{\kp'^2 + m^2}\\
& = -\frac{3}{16\pi}\frac{\kp^2 + 2m^2}{\kp^2 + 4m^2} - \frac{3}{16\pi}\frac{m^4}{k_\perp(\kp^2 + 4m^2)^{3/2}}\log\left[\frac{(\kp^2 + 2m^2)\left(\kp^2 + 2m^2- k_\perp\sqrt{\kp^2 + 4m^2}\right) - 2m^2}{(\kp^2 + 2m^2)\left(\kp^2 + 2m^2+ k_\perp\sqrt{\kp^2 + 4m^2}\right) - 2m^2}\right]\,.
\end{split}
\label{eq:kpprimeint}
\ee
For the final integration over $k_\perp$ we are only interested in extracting the leading log. We can drop the second term in \eqref{eq:kpprimeint} as it vanishes in the limit $m\to 0$. Focusing on the first term, we eventually find
\be
\begin{split}
&\frac{1}{2\pi}\int_0^{\Lambda}\frac{k_\perp \rmd k_\perp}{\kp^2 + m^2}\int_0^\infty \frac{ \calJ(k_\perp,k'_\perp) k'_\perp\rmd k'_\perp}{\kp'^2 + m^2} \simeq -\frac{3}{32\pi^2}\int_0^{\Lambda} \frac{k_\perp \rmd k_\perp}{\kp^2 + m^2}\frac{\kp^2 + 2m^2}{\kp^2 + 4m^2} \simeq -\frac{3}{32\pi^2}\log\left(\frac{1}{r_\perp m} + \rme\right)\,.
\end{split}
\ee
In total we have
\be
\Theta(\xp,\yp) = r_\perp^3 (\hat{\boldsymbol{r}}_\perp \cdot \hat{\boldsymbol{b}}_\perp) (\pi R_A^2)\frac{\rmd T_A(\bp)}{\rmd b_\perp} \frac{3}{32\pi^2}\log\left(\frac{1}{r_\perp m} + \rme\right)\,.
\ee
Using \eqref{eq:couplings} the prefactor in \eqref{eq:oddjv} is
\be
-g^3 C_{3F}\frac{\mu^6}{\kappa} = -\frac{\pi^2}{4}\frac{1}{\alpha_S^3}\frac{N_c^2 -4}{(N_c^2 - 1)^2}\frac{R_A^4}{A^2}Q_S^6\,.
\ee
Combining everything leads to
\be
\begin{split}
\calO(\rp,\bp) &= -\frac{3}{128}\frac{N_c^2 - 4}{(N_c^2 - 1)^2}\frac{Q_S^3 R_A^3}{\alpha_S^3 A^2}\left( R_A\frac{\rmd T_A(\bp)}{\rmd b_\perp}(\pi R_A^2)\right)(Q_S^3 r_\perp^3) (\hat{\boldsymbol{r}}_\perp \cdot \hat{\boldsymbol{b}}_\perp)\log\left(\frac{1}{r_\perp m} +  \rme\right)\\
&\times\exp\left[-\frac{1}{4}Q_S^2 \rp^2 (\pi R_A^2)T_A(\bp) \log\left(\frac{1}{r_\perp m}+\rme\right)\right]\,.
\end{split}
\label{eq:odd1}
\ee
A rather similar expression, that also involves the derivative of the transverse profile function was found in \cite{Kovchegov:2012ga}, see also \cite{Boer:2018vdi}. This expression is usually found in terms of a single transverse coordinate integral that can be solved \cite{Zhou:2013gsa} to get the $\calO(\rp,\bp)\sim r_\perp^3$ behavior.

\bibliographystyle{h-physrev}
\bibliography{references}

\begin{thebibliography}{10}

\bibitem{Lukaszuk:1973nt}
L.~Lukaszuk and B.~Nicolescu,
\newblock Lett. Nuovo Cim. {\bf 8}, 405 (1973).

\bibitem{Ewerz:2003xi}
C.~Ewerz,
\newblock (2003), hep-ph/0306137.

\bibitem{Donnachie:1985iz}
A.~Donnachie and P.~V. Landshoff,
\newblock Nucl. Phys. B {\bf 267}, 690 (1986).

\bibitem{TOTEM:2020zzr}
TOTEM, D0, V.~M. Abazov {\em et~al.},
\newblock Phys. Rev. Lett. {\bf 127}, 062003 (2021), 2012.03981.

\bibitem{TOTEM:2018psk}
TOTEM, G.~Antchev {\em et~al.},
\newblock Eur. Phys. J. C {\bf 80}, 91 (2020), 1812.08610.

\bibitem{D0:2012erd}
D0, V.~M. Abazov {\em et~al.},
\newblock Phys. Rev. D {\bf 86}, 012009 (2012), 1206.0687.

\bibitem{Czyzewski:1996bv}
J.~Czyzewski, J.~Kwiecinski, L.~Motyka, and M.~Sadzikowski,
\newblock Phys. Lett. B {\bf 398}, 400 (1997), hep-ph/9611225,
\newblock [Erratum: Phys.Lett.B 411, 402 (1997)].

\bibitem{Engel:1997cga}
R.~Engel, D.~Y. Ivanov, R.~Kirschner, and L.~Szymanowski,
\newblock Eur. Phys. J. C {\bf 4}, 93 (1998), hep-ph/9707362.

\bibitem{Bartels:2001hw}
J.~Bartels, M.~A. Braun, D.~Colferai, and G.~P. Vacca,
\newblock Eur. Phys. J. C {\bf 20}, 323 (2001), hep-ph/0102221.

\bibitem{Ma:2003py}
J.~P. Ma,
\newblock Nucl. Phys. A {\bf 727}, 333 (2003), hep-ph/0301155.

\bibitem{Goncalves:2012cy}
V.~P. Goncalves,
\newblock Nucl. Phys. A {\bf 902}, 32 (2013), 1211.1207.

\bibitem{Goncalves:2015hra}
V.~P. Goncalves and W.~K. Sauter,
\newblock Phys. Rev. D {\bf 91}, 094014 (2015), 1503.05112.

\bibitem{Goncalves:2018yxc}
V.~P. Gon\c{c}alves and B.~D. Moreira,
\newblock Phys. Rev. D {\bf 97}, 094009 (2018), 1801.10501.

\bibitem{Harland-Lang:2018ytk}
L.~A. Harland-Lang, V.~A. Khoze, A.~D. Martin, and M.~G. Ryskin,
\newblock Phys. Rev. D {\bf 99}, 034011 (2019), 1811.12705.

\bibitem{Dumitru:2019qec}
A.~Dumitru and T.~Stebel,
\newblock Phys. Rev. D {\bf 99}, 094038 (2019), 1903.07660.

\bibitem{Babiarz:2023cac}
I.~Babiarz, V.~P. Goncalves, W.~Sch\"afer, and A.~Szczurek,
\newblock (2023), 2306.00754.

\bibitem{Accardi:2012qut}
A.~Accardi {\em et~al.},
\newblock Eur. Phys. J. A {\bf 52}, 268 (2016), 1212.1701.

\bibitem{LHeCStudyGroup:2012zhm}
LHeC Study Group, J.~L. Abelleira~Fernandez {\em et~al.},
\newblock J. Phys. G {\bf 39}, 075001 (2012), 1206.2913.

\bibitem{Anderle:2021wcy}
D.~P. Anderle {\em et~al.},
\newblock Front. Phys. (Beijing) {\bf 16}, 64701 (2021), 2102.09222.

\bibitem{Bertulani:2005ru}
C.~A. Bertulani, S.~R. Klein, and J.~Nystrand,
\newblock Ann. Rev. Nucl. Part. Sci. {\bf 55}, 271 (2005), nucl-ex/0502005.

\bibitem{Iancu:2003xm}
E.~Iancu and R.~Venugopalan,
\newblock {The Color glass condensate and high-energy scattering in QCD},
\newblock in {\em {Quark-gluon plasma 3}}, edited by R.~C. Hwa and X.-N. Wang,
  pp. 249--3363, World Scientific, 2003, hep-ph/0303204.

\bibitem{Gelis:2010nm}
F.~Gelis, E.~Iancu, J.~Jalilian-Marian, and R.~Venugopalan,
\newblock Ann. Rev. Nucl. Part. Sci. {\bf 60}, 463 (2010), 1002.0333.

\bibitem{Kovchegov:2012mbw}
Y.~V. Kovchegov and E.~Levin,
\newblock {Quantum chromodynamics at high energy}, 2012.

\bibitem{Blaizot:2016qgz}
J.-P. Blaizot,
\newblock Rept. Prog. Phys. {\bf 80}, 032301 (2017), 1607.04448.

\bibitem{Kovchegov:2003dm}
Y.~V. Kovchegov, L.~Szymanowski, and S.~Wallon,
\newblock Phys. Lett. B {\bf 586}, 267 (2004), hep-ph/0309281.

\bibitem{Hatta:2005as}
Y.~Hatta, E.~Iancu, K.~Itakura, and L.~McLerran,
\newblock Nucl. Phys. A {\bf 760}, 172 (2005), hep-ph/0501171.

\bibitem{Motyka:2005ep}
L.~Motyka,
\newblock Phys. Lett. B {\bf 637}, 185 (2006), hep-ph/0509270.

\bibitem{Lappi:2016gqe}
T.~Lappi, A.~Ramnath, K.~Rummukainen, and H.~Weigert,
\newblock Phys. Rev. D {\bf 94}, 054014 (2016), 1606.00551.

\bibitem{Yao:2018vcg}
X.~Yao, Y.~Hagiwara, and Y.~Hatta,
\newblock Phys. Lett. B {\bf 790}, 361 (2019), 1812.03959.

\bibitem{Lappi:2013zma}
T.~Lappi and H.~M\"antysaari,
\newblock Phys. Rev. D {\bf 88}, 114020 (2013), 1309.6963.

\bibitem{Dumitru:2022ooz}
A.~Dumitru, H.~M\"antysaari, and R.~Paatelainen,
\newblock Phys. Rev. D {\bf 107}, L011501 (2023), 2210.05390.

\bibitem{Jeon:2005cf}
S.~Jeon and R.~Venugopalan,
\newblock Phys. Rev. D {\bf 71}, 125003 (2005), hep-ph/0503219.

\bibitem{Jalilian-Marian:2005ccm}
J.~Jalilian-Marian and Y.~V. Kovchegov,
\newblock Prog. Part. Nucl. Phys. {\bf 56}, 104 (2006), hep-ph/0505052.

\bibitem{Mantysaari:2020lhf}
H.~M\"antysaari, K.~Roy, F.~Salazar, and B.~Schenke,
\newblock Phys. Rev. D {\bf 103}, 094026 (2021), 2011.02464.

\bibitem{Kowalski:2006hc}
H.~Kowalski, L.~Motyka, and G.~Watt,
\newblock Phys. Rev. D {\bf 74}, 074016 (2006), hep-ph/0606272.

\bibitem{McLerran:1998nk}
L.~D. McLerran and R.~Venugopalan,
\newblock Phys. Rev. D {\bf 59}, 094002 (1999), hep-ph/9809427.

\bibitem{Balitsky:2001mr}
I.~I. Balitsky and A.~V. Belitsky,
\newblock Nucl. Phys. B {\bf 629}, 290 (2002), hep-ph/0110158.

\bibitem{Blaizot:2004wv}
J.~P. Blaizot, F.~Gelis, and R.~Venugopalan,
\newblock Nucl. Phys. A {\bf 743}, 57 (2004), hep-ph/0402257.

\bibitem{Boussarie:2019vmk}
R.~Boussarie, Y.~Hatta, L.~Szymanowski, and S.~Wallon,
\newblock Phys. Rev. Lett. {\bf 124}, 172501 (2020), 1912.08182.

\bibitem{Hatta:2017cte}
Y.~Hatta, B.-W. Xiao, and F.~Yuan,
\newblock Phys. Rev. D {\bf 95}, 114026 (2017), 1703.02085.

\bibitem{Lappi:2020ufv}
T.~Lappi, H.~M\"antysaari, and J.~Penttala,
\newblock Phys. Rev. D {\bf 102}, 054020 (2020), 2006.02830.

\bibitem{Dosch:1996ss}
H.~G. Dosch, T.~Gousset, G.~Kulzinger, and H.~J. Pirner,
\newblock Phys. Rev. D {\bf 55}, 2602 (1997), hep-ph/9608203.

\bibitem{Pham:2007xx}
T.~N. Pham,
\newblock AIP Conf. Proc. {\bf 964}, 124 (2007), 0710.2846.

\bibitem{Baltz:2001dp}
A.~J. Baltz, F.~Gelis, L.~D. McLerran, and A.~Peshier,
\newblock Nucl. Phys. A {\bf 695}, 395 (2001), nucl-th/0101024.

\bibitem{Gelis:2001da}
F.~Gelis and A.~Peshier,
\newblock Nucl. Phys. A {\bf 697}, 879 (2002), hep-ph/0107142.

\bibitem{Gelis:2001dh}
F.~Gelis and A.~Peshier,
\newblock Nucl. Phys. A {\bf 707}, 175 (2002), hep-ph/0111227.

\bibitem{Peskin:1995ev}
M.~E. Peskin and D.~V. Schroeder,
\newblock {\em {An Introduction to quantum field theory}} (Addison-Wesley,
  Reading, USA, 1995).

\bibitem{Ye:2017gyb}
Z.~Ye, J.~Arrington, R.~J. Hill, and G.~Lee,
\newblock Phys. Lett. B {\bf 777}, 8 (2018), 1707.09063.

\bibitem{STAR:2022wfe}
STAR, M.~Abdallah {\em et~al.},
\newblock Sci. Adv. {\bf 9}, eabq3903 (2023), 2204.01625.

\bibitem{Mantysaari:2022sux}
H.~M\"antysaari, F.~Salazar, and B.~Schenke,
\newblock Phys. Rev. D {\bf 106}, 074019 (2022), 2207.03712.

\bibitem{Balitsky:1995ub}
I.~Balitsky,
\newblock Nucl. Phys. B {\bf 463}, 99 (1996), hep-ph/9509348.

\bibitem{Kovchegov:1999yj}
Y.~V. Kovchegov,
\newblock Phys. Rev. D {\bf 60}, 034008 (1999), hep-ph/9901281.

\bibitem{Golec-Biernat:2003naj}
K.~J. Golec-Biernat and A.~M. Stasto,
\newblock Nucl. Phys. B {\bf 668}, 345 (2003), hep-ph/0306279.

\bibitem{Ikeda:2004zp}
T.~Ikeda and L.~McLerran,
\newblock Nucl. Phys. A {\bf 756}, 385 (2005), hep-ph/0410345.

\bibitem{Berger:2010sh}
J.~Berger and A.~Stasto,
\newblock Phys. Rev. D {\bf 83}, 034015 (2011), 1010.0671.

\bibitem{Hagiwara:2016kam}
Y.~Hagiwara, Y.~Hatta, and T.~Ueda,
\newblock Phys. Rev. D {\bf 94}, 094036 (2016), 1609.05773.

\bibitem{Cepila:2018faq}
J.~Cepila, J.~G. Contreras, and M.~Matas,
\newblock Phys. Rev. D {\bf 99}, 051502 (2019), 1812.02548.

\bibitem{Kowalski:2008sa}
H.~Kowalski, T.~Lappi, C.~Marquet, and R.~Venugopalan,
\newblock Phys. Rev. C {\bf 78}, 045201 (2008), 0805.4071.

\bibitem{Balitsky:2006wa}
I.~Balitsky,
\newblock Phys. Rev. D {\bf 75}, 014001 (2007), hep-ph/0609105.

\bibitem{Braun:2020vmd}
M.~A. Braun,
\newblock Phys. Lett. B {\bf 809}, 135742 (2020), 2005.11049.

\bibitem{Contreras:2020lrh}
C.~Contreras, E.~Levin, R.~Meneses, and M.~Sanhueza,
\newblock Phys. Rev. D {\bf 101}, 096019 (2020), 2004.04445.

\bibitem{DeVries:1987atn}
H.~De~Vries, C.~W. De~Jager, and C.~De~Vries,
\newblock Atom. Data Nucl. Data Tabl. {\bf 36}, 495 (1987).

\bibitem{GENZ1980}
A.~Genz and A.~Malik,
\newblock Journal of Computational and Applied Mathematics {\bf 6}, 295 (1980).

\bibitem{GENZ1991}
J.~Berntsen, T.~O. Espelid, and A.~Genz,
\newblock ACM Trans. Math. Softw. {\bf 17}, 437–451 (1991).

\bibitem{Kaiser_2006}
N.~Kaiser,
\newblock Journal of Physics A: Mathematical and General {\bf 39}, 15287
  (2006).

\bibitem{Kang:2019ctl}
Z.-B. Kang, A.~Prokudin, N.~Sato, and J.~Terry,
\newblock Comput. Phys. Commun. {\bf 258}, 107611 (2021), 1906.05949.

\bibitem{Dumitru:2020fdh}
A.~Dumitru, V.~Skokov, and T.~Stebel,
\newblock Phys. Rev. D {\bf 101}, 054004 (2020), 2001.04516.

\bibitem{CLAS:2011qvj}
CLAS, N.~Baillie {\em et~al.},
\newblock Phys. Rev. Lett. {\bf 108}, 142001 (2012), 1110.2770,
\newblock [Erratum: Phys.Rev.Lett. 108, 199902 (2012)].

\bibitem{Friscic:2021oti}
I.~Friscic {\em et~al.},
\newblock Phys. Lett. B {\bf 823}, 136726 (2021), 2106.08805.

\bibitem{Klein:2019qfb}
S.~R. Klein and H.~M\"antysaari,
\newblock Nature Rev. Phys. {\bf 1}, 662 (2019), 1910.10858.

\bibitem{AbdulKhalek:2021gbh}
R.~Abdul~Khalek {\em et~al.},
\newblock Nucl. Phys. A {\bf 1026}, 122447 (2022), 2103.05419.

\bibitem{Workman:2022ynf}
Particle Data Group, R.~L. Workman {\em et~al.},
\newblock PTEP {\bf 2022}, 083C01 (2022).

\bibitem{Klein:2018ypk}
S.~R. Klein,
\newblock Phys. Rev. D {\bf 98}, 118501 (2018), 1808.08253.

\bibitem{BaBar:2010siw}
BaBar, J.~P. Lees {\em et~al.},
\newblock Phys. Rev. D {\bf 81}, 052010 (2010), 1002.3000.

\bibitem{Dumitru:2011zz}
A.~Dumitru, J.~Jalilian-Marian, and E.~Petreska,
\newblock Phys. Rev. D {\bf 84}, 014018 (2011), 1105.4155.

\bibitem{Dumitru:2018vpr}
A.~Dumitru, G.~A. Miller, and R.~Venugopalan,
\newblock Phys. Rev. D {\bf 98}, 094004 (2018), 1808.02501.

\bibitem{Kovchegov:2012ga}
Y.~V. Kovchegov and M.~D. Sievert,
\newblock Phys. Rev. D {\bf 86}, 034028 (2012), 1201.5890,
\newblock [Erratum: Phys.Rev.D 86, 079906 (2012)].

\bibitem{Boer:2018vdi}
D.~Boer, T.~Van~Daal, P.~J. Mulders, and E.~Petreska,
\newblock JHEP {\bf 07}, 140 (2018), 1805.05219.

\bibitem{Zhou:2013gsa}
J.~Zhou,
\newblock Phys. Rev. D {\bf 89}, 074050 (2014), 1308.5912.

\end{thebibliography}

\end{document}